\pdfoutput=1
\documentclass[nofootinbib,aps]{revtex4}
\usepackage{amssymb,amsmath,graphicx,color,microtype}
\usepackage{enumerate}
\usepackage{slashed}
\usepackage{hyperref}
\hypersetup{
	colorlinks=true,
	linkcolor=blue,
	filecolor=blue,      
	urlcolor=blue,
	pdftitle={Sharelatex Example},
	bookmarks=true,
	citecolor=blue
}
\usepackage[paperwidth=210mm,paperheight=297mm,centering,hmargin=2cm,vmargin=2.5cm]{geometry}

\begin{document}
\title{Connections between Minkowski and Cosmological Correlation Functions} 
	\author{Shek Kit Chu$^{1,2}$}
	\author{Mang Hei Gordon Lee$^{1}$}
	\author{Shiyun Lu$^{1,2}$}
    \author{Xi Tong$^{3}$}
	\author{Yi Wang$^{1,2}$}
	\author{Siyi Zhou$^{1,2}$ \vspace{3mm}}
	\email{skchuab@connect.ust.hk, mhglee@connect.ust.hk, sluah@connect.ust.hk, tx123@mail.ustc.edu.cn, phyw@ust.hk, szhouah@connect.ust.hk}
	
	\affiliation{${}^1$Department of Physics, The Hong Kong University of Science and Technology, \\
		Clear Water Bay, Kowloon, Hong Kong, P.R.China}
	\affiliation{${}^2$Jockey Club Institute for Advanced Study, The Hong Kong University of Science and Technology, \\
		Clear Water Bay, Kowloon, Hong Kong, P.R.China}
	\affiliation{${}^3$School of Physics, University of Science and Technology of China, \\
	Hefei, Anhui 230026, China}
	
\begin{abstract}
We show how cosmological correlation functions of massless fields can be rewritten in terms of Minkowski correlation functions, by extracting symmetry-breaking operators from the cosmological correlators. This technique simplifies some cosmological calculations. Also, known properties of Minkowski correlation functions can be translated to non-trivial properties of cosmological correlations. To illustrate this idea, inflation to Minkowski and matter bounce to Minkowski relations are presented for the interactions of general single field inflation. And a Minkowski recursion relation is translated to a novel relation for inflation.
\end{abstract}
	
\maketitle

\section{Introduction}\label{introduction}	
Symmetry plays an important role in understanding correlation functions in quantum field theory. In realistic problems, some symmetries may be broken (either hard or spontaneously), in which case we lose the ability to directly understand the correlations purely in terms of the broken symmetries. For instance, in cosmology, Lorentz invariance and time translation are spontaneously broken by the time-dependent background. As a result, some of the features of cosmological correlation functions related to spacetime symmetry become opaque. However, it is sometimes possible to rewrite an asymmetric correlation function into
\begin{align}
\Big\langle \rm asymmetric\,\, correlation\,\, function \Big\rangle' 
= (asymmetric\,\, operator) \times \Big\langle symmetric\,\, correlation\,\, function \Big\rangle',
\end{align}
where $\langle\cdots\rangle'$ denotes correlation functions with the momentum conservation delta function stripped.
Such a possibility enables us to extend the usage of symmetry and its consequences to asymmetric correlation functions. In this paper, we provide explicit examples of correlation functions in Friedmann-Robertson-Walker (FRW) cosmologies, which can be obtained by applying asymmetric operators on correlation functions with the symmetry of Minkowski space.

It was noted~\cite{Arkani-Hamed:2015bza} that in de Sitter space, correlation functions of massless scalars with exchange of massive scalars~\cite{Chen:2009we,Chen:2009zp,Baumann:2011nk,Assassi:2012zq,Noumi:2012vr} can be obtained by applying a differential operator on correlation functions with conformally-coupled scalars. The relation takes the form:
\begin{align}
  \Big \langle \mbox{massless scalars} \Big \rangle'_\mathrm{inflation}
  =
 {O} \Big \langle \mbox{conformal scalars} \Big \rangle'_\mathrm{inflation}~,
\end{align}
where the operator $O$ is constructed by rewriting the time variables into $\partial_K$ for some combination of three-momentum $K$ and then pull this operator out of the time integration of in-in formalism. This relation connects different types of field for inflationary (approximately de Sitter) spacetime, and should be understood in a diagram-by-diagram sense. 

In this paper, we extend this observation to relate the correlation functions of massless fields in FRW (inflation or matter bounce) and Minkowski backgrounds. For inflation, the relation between inflationary and flat-space correlators takes a general form:
\begin{align}
  \Big \langle \mbox{curvature perturbation} \Big \rangle'_\mathrm{inflation}
  =
  \sum_i {O}_i \Big \langle \mbox{massless scalars} \Big \rangle'_{\mathrm{flat}, i}~,
\end{align}
where the index $i$ stands for different subprocesses with different contractions.

Interestingly, for some specific types of interactions such as a simple $\dot\zeta^3$ vertex (where $\zeta$ is the curvature fluctuation in comoving gauge), the operators corresponding to different contraction coincides with each other. This yields a stronger relation that we are able to pull out an overall operator acting on the Minkowski correlation to generate the correlation in cosmology:
\begin{align}
  \Big \langle \mbox{curvature perturbation} \Big \rangle'_\mathrm{inflation}
  =
   O \Big \langle \mbox{massless scalars} \Big \rangle'_{\mathrm{flat}}~.
\end{align}
This type of method can be used to compute correlation functions in other kinds of cosmological background as well, such as the matter bounce cosmology. 

With the relation between cosmological and Minkowski correlation functions, one can import known relations of Minkowski correlation functions to cosmology. For example, in~\cite{Arkani-Hamed:2017fdk}, a BCFW-like recursion relation~\cite{Britto:2005fq,ArkaniHamed:2010kv,ArkaniHamed:2012nw,Benincasa:2015zna} is derived for the wave function of the universe for conformally-coupled scalars. By applying the operator technique, we are able to obtain the corresponding recursion relations for massless fields in cosmology.  

This paper is organized as follows. In Section~\ref{generalmethod}, we compute the equal-time correlation functions of massless scalars in flat space, and introduce a general formalism can be used to relate these correlators to their cosmological counterpart. In Section~\ref{applicationtoinflation}, we apply this formalism to general single-field inflation. In Section~\ref{recursionrelation}, we apply our techniques to obtain a recursion relation between the cosmological correlation functions by using the Minkowski recursion relation. We conclude in Section~\ref{conclusionandoutlook}.

\section{General Method}\label{generalmethod}

In this section, we rewrite the Minkowski correlators into the three-dimensional forms which are similar to FRW correlators, and outline how the relations between Minkowski and FRW is to be constructed. In \S\ref{flat}, we compute equal-time three- and four-point functions in flat space. We describe the rule to translate these into inflationary correlation functions in \S\ref{transitiontocosmology}.
  
\subsection{Minkowski Correlator}\label{flat}
We start by writing down some equal-time correlators in Minkowski spacetime, these correlators will be later used to generate equal-time correlators in inflation and bouncing cosmologies. Consider the action 
\begin{align}
	S_0  = - \frac{1}{2}\int d\tau d^3 x\,  (\partial\phi)^2 ~.
\end{align}
Here the $(-,+,+,+)$ metric convention is used. The field $\phi$ is quantized as
\begin{align}
	\phi_{\mathbf k} (\tau) = u_k (\tau) a_{\mathbf k} + u^*_k (\tau) a^\dagger_{-\mathbf k}~,
\end{align}
where $a_{\mathbf k}$ and $a^\dagger_{-\mathbf k}$ are the creation and annihilation operators satisfying the canonical commutation relations.
The mode function is given by 
\begin{align}\label{flatuk}
	u_k(\tau) = \frac{1}{\sqrt{2k}} e^{-i k\tau}\, .
\end{align}
As examples, we consider two kinds of interaction Hamiltonians of the free field,
\begin{align}
H_{\phi^3} =  \int d^3 x\, \lambda_3\,\phi^3\, ,  \quad H_{\phi^4} =  \int d^3 x\, \lambda_4\,\phi^4\, ,
\end{align}
where $\lambda_3$ and $\lambda_4$ are coupling constants, and use the in-in formalism to calculate the correlation in the Minkowski spacetime with $H_{I} = H_{\phi^3} + H_{\phi^4}$.
The equal-time $n$-point function of $\phi$ can then be computed using the in-in formalism~\cite{Weinberg:2005vy,Chen:2010xka,Wang:2013eqj}
\begin{align}\label{expression1}
\langle\phi^n(\tau)\rangle  = \Big\langle \Big[ \bar T e^{i \int_{-\infty}^0d \tau'\, H_{I} (\tau')} \Big]\, \phi^n(\tau)\, \Big[ T e^{-i \int_{-\infty}^0d\tau'\, H_{I} (\tau') } \Big] \Big\rangle\, ,
\end{align}
where $T$ and $\bar T$ denote the time-ordering and anti-time-ordering operators, respectively.

\subsubsection{Three-Point Function}
\begin{figure}[htbp] 
	\centering 
	\includegraphics[width=3cm]{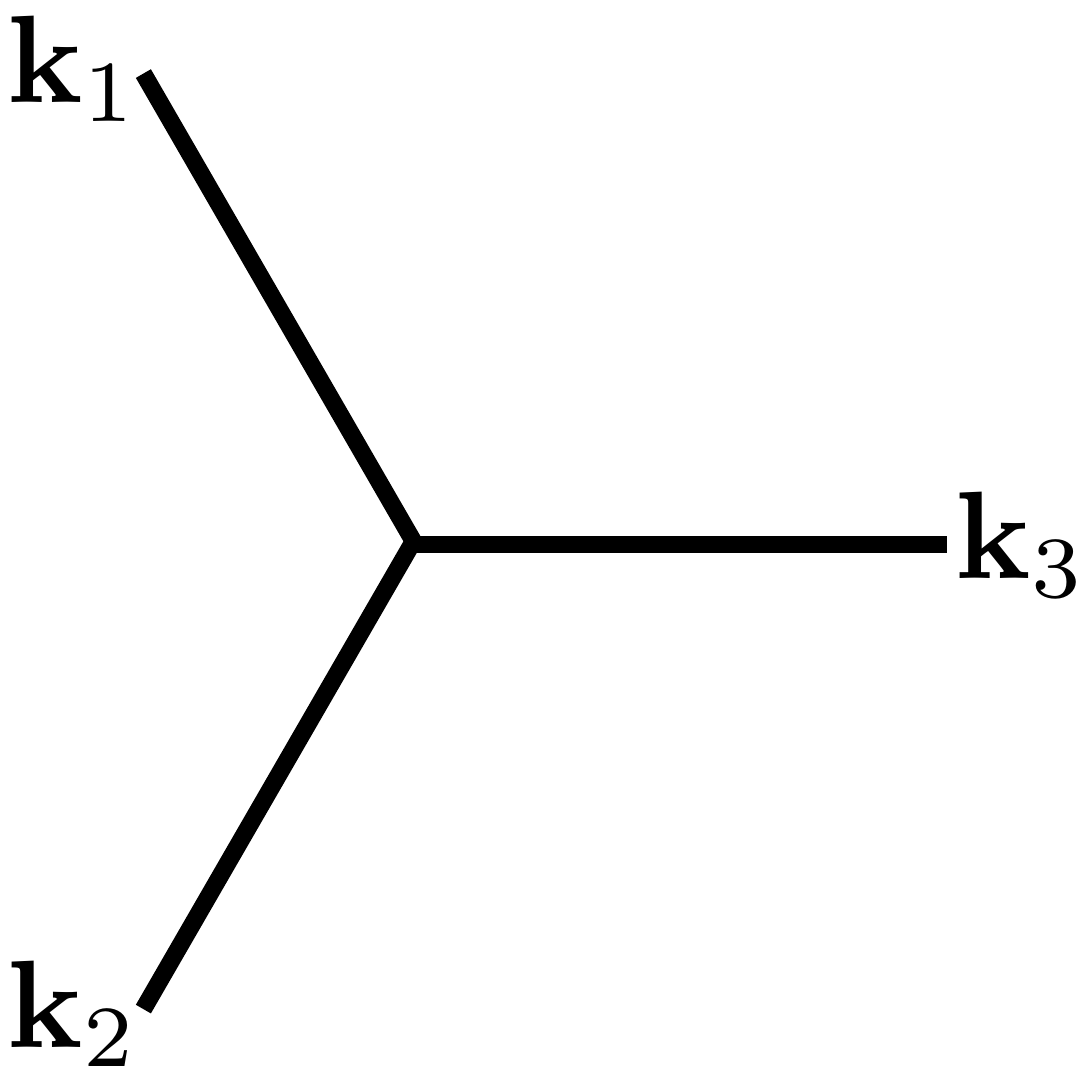}  
	\caption{\label{fig:3pt} The Feynman diagram associated to the three-point function.} 
\end{figure}

The three-point function involves only the $H_{\phi^3}$ interaction at tree level. The leading-order correlation function is calculated from the first order contribution in the perturbation series
\begin{align}
	\langle \phi_{\mathbf k_1}\phi_{\mathbf k_2}\phi_{\mathbf k_3} \rangle = 2 {\rm Im} \int_{\tau_0}^{\tau} d \tau_1 \langle 0 | \phi_{\mathbf k_1}\phi_{\mathbf k_2}\phi_{\mathbf k_3} H_{\phi^3}(\tau_1) | 0 \rangle~.
\end{align} 
The corresponding Feynman diagram is illustrated in Fig.~\ref{fig:3pt}.
Taking the initial time to $-\infty$ and final time to $0$, we get
\begin{align}
	\langle \phi_{\mathbf k_1}\phi_{\mathbf k_2}\phi_{\mathbf k_3} \rangle' = -\frac{3\lambda_3}{2k_1k_2k_3k_{123}}\, ,
\end{align}
where we use the notation $k_{i_1\cdots i_n}\equiv k_{i_1}+\cdots+k_{i_n}$, and the prime indicates that we strip the momentum conserving delta function $(2\pi)^3 \delta (\mathbf k_1+\mathbf k_2 + \mathbf k_3)$. 

\subsubsection{Four-Point Function}
\begin{figure}[htbp] 
	\centering 
	\includegraphics[width=10cm]{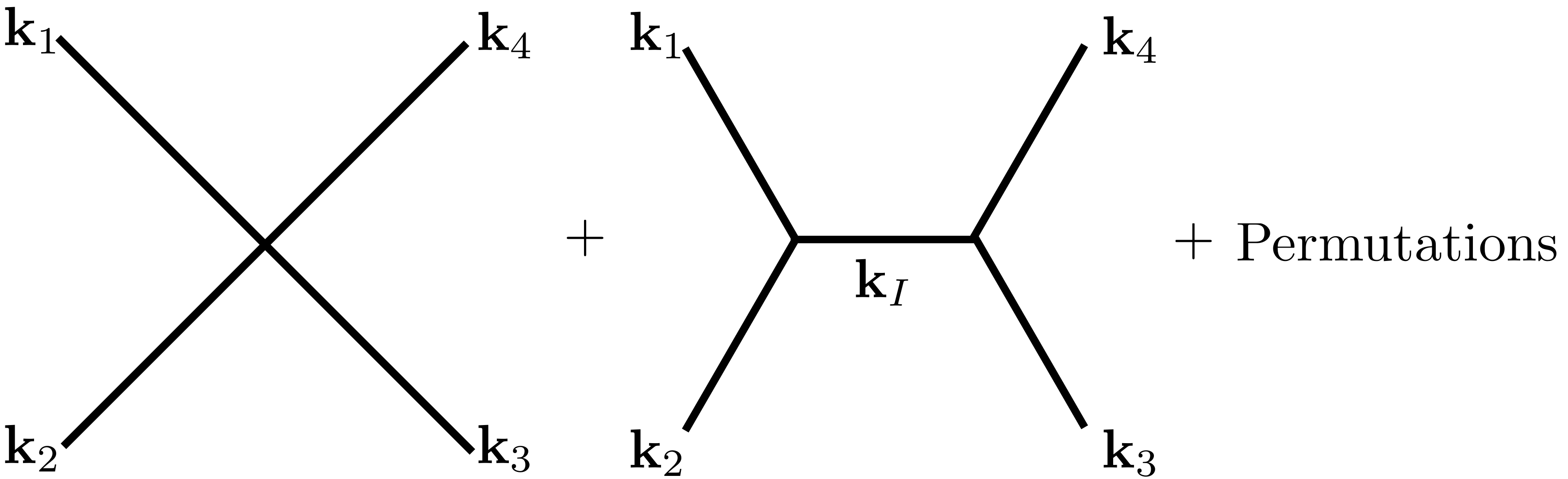}   
	\caption{\label{fig:4pt}The Feynman diagrams associated to the four-point function.} 
\end{figure}

The four-point function involves both $H_{\phi^3}$ and $H_{\phi^4}$ interactions. The leading order correlation function is calculated from the first order and the second order in the perturbation series (as shown in Fig.~\ref{fig:4pt}), respectively:
\begin{align}\nonumber
	\langle \phi_{\mathbf k_1}\phi_{\mathbf k_2}\phi_{\mathbf k_3}\phi_{\mathbf k_4} \rangle & = 2 {\rm Im} \int_{\tau_0}^{\tau} d \tau_1 \langle 0 | \phi_{\mathbf k_1}\phi_{\mathbf k_2}\phi_{\mathbf k_3}\phi_{\mathbf k_4} H_{\phi^4}(\tau_1) | 0 \rangle \\ \nonumber
	& + \int_{\tau_0}^{\tau} d\tau_1 \int_{\tau_0}^{\tau} d\tau_2 \langle 0 | H_{\phi^3} (\tau_1) \phi_{\mathbf k_1}\phi_{\mathbf k_2}\phi_{\mathbf k_3}\phi_{\mathbf k_4} H_{\phi^3} (\tau_2) | 0 \rangle \\
	& - 2 {\rm Re} \int_{\tau_0}^{\tau} d\tau_1 \int_{\tau_0}^{\tau_1} d\tau_2 \langle 0 |  \phi_{\mathbf k_1}\phi_{\mathbf k_2}\phi_{\mathbf k_3}\phi_{\mathbf k_4} H_{\phi^3} (\tau_1) H_{\phi^3} (\tau_2) | 0 \rangle~.
\end{align}
Taking the initial time going to $-\infty$ and final time going to $0$, we get
\begin{align}
	\langle \phi_{\mathbf k_1}\phi_{\mathbf k_2}\phi_{\mathbf k_3}\phi_{\mathbf k_4} \rangle' = \langle \phi_{\mathbf k_1}\phi_{\mathbf k_2}\phi_{\mathbf k_3}\phi_{\mathbf k_4} \rangle'_{(3)} + \langle \phi_{\mathbf k_1}\phi_{\mathbf k_2}\phi_{\mathbf k_3}\phi_{\mathbf k_4} \rangle'_{(4)}~,
\end{align}
where the subscripts $(3)$ and $(4)$ denote the diagrams contributed by the three-point interaction and four-point interaction, respectively. The part of contributions from three-point interaction is
\begin{align}
\langle \phi_{\mathbf k_1}\phi_{\mathbf k_2}\phi_{\mathbf k_3}\phi_{\mathbf k_4} \rangle'_{(3)}	& =  \frac{9\lambda_3^2}{2 k_1 k_2 k_3 k_4 k_t}\frac{k_t+k_I}{k_I(k_{12}+k_I)(k_{34}+k_I)} + \text{permutations} \, .
\end{align}
where we denoted the total momentum by $k_t=k_{1234}$ and the internal momentum by $k_I=|\mathbf k_1+\mathbf k_2|$. 
The contact diagram gives
\begin{align}
\langle \phi_{\mathbf k_1}\phi_{\mathbf k_2}\phi_{\mathbf k_3}\phi_{\mathbf k_4} \rangle'_{(4)}=- \frac{3\lambda_4}{k_1k_2k_3k_4k_t}\, . 
\end{align}

\subsection{From Minkowski to FRW}\label{transitiontocosmology}

Here we sketch the relation between Minkowski and cosmological correlation functions for a massless field. Here we use inflation as an example (where details will be described in Section~\ref{applicationtoinflation} and Appendix \ref{appendA}). The method can be readily generalized to matter dominated contracting universe (where details will be described in Appendix~\ref{applicationtobounce}), as well as other cosmologies where the solution of the massless field can be written in terms of elementary functions (such as matter dominated expansion, or radiation dominated contracting or expanding universes).

The inflation mode functions at $\tau=0$ can be related to Minkowski spacetime by
\begin{align}
u_k(0)_{\rm inflation} = C u_k(0)_{\rm flat}~.
\end{align}
The Hamiltonian is then related to the Minkowski Hamiltonian by 
\begin{align}
\int d\tau H_I(\tau)_{\rm inflation} \rightarrow O_{\mathrm{ext},i} \int d\tau H_I(\tau)_{\mathrm{ flat}} ~,
\end{align}
where $i$ denotes different contractions in the in-in formalism. The subscript ``ext'' means that usually this operator involves integration or derivative with respect to the total momentum of the external legs associated to a certain vertex. Such a relation is possible because the inflationary mode function contains oscillatory parts and the Hamiltonian in general takes the form $\int d\tau f(\tau) e^{\pm i K \tau}$, where $K$ is a combination of external momenta to be specified in the next section. As a result, the $f(\tau)$ part of the integrand can be replaced by $f(\mp i \partial_K)$ and be extracted out of the integral. The remaining part is the Minkowski Hamiltonian up to constant normalization factors.

Putting them altogether, we have
\begin{align}
\langle \cdots \rangle'_{\mathrm{inflation},i} = O_{i} \langle \cdots \rangle'_{\mathrm{flat},i} ~.
\end{align}
$O_{i}$ is related to $O_{\mathrm{ext},i}$ by 
\begin{align}
O_{i} = O_{\mathrm{ext},i}^m ~ C^n~,
\end{align}
for $n$-point function and $m$-th order in Hamiltonian.  Note that this is a schematic relation. For different Hamiltonians the $O_{\mathrm{ext},i}$ operators are different, and the $C$ factors depends on the values of external momenta.

For some interactions (for example $\dot\zeta^3$ as we will show in the next section), the operators associated with different types of in-in contours degenerate, so that we can arrive at a stronger statement
\begin{align}
\langle \cdots \rangle'_{\rm inflation} = O \langle \cdots \rangle'_{\rm flat} ~,
\end{align}
where $O$ is related to $O_{\rm ext}$ by
\begin{align}
O = O_{\rm ext}^m ~ C^n~.
\end{align}
For different Hamiltonians the $O_{\rm ext}$ operators are different, but for different in-in contours these operators are the same for such specific interactions. 

\section{Application to Inflation}\label{applicationtoinflation}
In this section, we would like to use the symmetry-breaking operator technique to calculate the three-point function in general single field inflation~\cite{Chen:2006nt} as an illustration\footnote{From the perspective of symmetry, there are other efforts along the line of the dS/CFT correspondence~\cite{Strominger:2001pn,Strominger:2001gp} which makes the symmetry structure of inflationary correlation functions manifest~\cite{McFadden:2010na,McFadden:2010vh,Isono:2016yyj}. A nearly Minkowski spacetime with slow expansion can also be related to inflation by a cosmic duality~\cite{Piao:2011bz}.}.

The three-point function generated by single field inflation models with a canonical kinetic term is suppressed by slow-roll parameters~\cite{Maldacena:2002vr,Acquaviva:2002ud}. Interactions in the generalized Lagrangians~\cite{ArmendarizPicon:1999rj,Alishahiha:2004eh} are known to source potentially large three-point functions~\cite{Chen:2006nt}. Here we will apply the symmetry-breaking operator method to this model. The readers not interested in the technical details can directly find the resulting inflation-Minkowski relation for three-point functions in Equations \eqref{eq:3pt3a} and \eqref{eq:3ptdzpzpz}. We will also derive the relation for four-point functions in Appendix~\ref{appendA}.

In the following we set the Planck mass $M_{\rm pl}$ to 1. The Lagrangian for general single field inflation is
\begin{align}
	S = \frac{1}{2} \int d^4 x \sqrt{-g} \bigg[ R + 2 P(X,\phi)  \bigg]~,
\end{align}
where $\phi$ is the inflaton field and $X = -\frac{1}{2}g^{\mu\nu} \partial_\mu\phi\partial_\nu\phi$. The three slow-variation parameters are defined as 
\begin{align}
	\epsilon = - \frac{\dot H}{H^2}, \quad \eta = \frac{\dot \epsilon}{\epsilon H}, \quad s = \frac{\dot c_s}{c_s H}~, 
\end{align}
where $H=\dot a/a$ is the Hubble parameter and the sound speed is 
\begin{align}
	c_s^2\equiv \frac{P_{,X}}{P_{,X}+ 2 X P_{,XX}}~,
\end{align}
where $P_{,X}$ denote the derivative of $P(X,\phi)$ with respect to $X$.
There are two other useful parameters $\lambda$ and $\Sigma$ defined as~\cite{Seery:2005wm}
\begin{align}
	\lambda & =  X^2 P_{,XX} + \frac{2}{3} X^3 P_{,XX}~, \\
	\Sigma &  = X P_{,X} + 2 X^2 P_{,XX}~.
\end{align} 
We will focus on cases with either $c_s \ll 1$ or $|\lambda/\Sigma| \gg 1$, such that the corresponding interactions dominate over the slow roll contributions. The method can be straightforwardly generalized to these slow roll suppressed terms.

The second order action for general single field inflation is
\begin{align}
	S_2 = \int dt d^3 x \bigg[ a^3 \frac{\epsilon}{c_s^2} \dot{\zeta}^2 - a\epsilon (\partial \zeta)^2  \bigg]~,
\end{align}
where $\zeta$ is quantized as
\begin{align}
	\zeta_{\mathbf k} (\tau) = u_k (\tau) a_{\mathbf k} + u^*_k (\tau) a^\dagger_{-\mathbf k}~,
\end{align}
with the usual commutation relations for the creation and annihilation operators. The mode function and its time derivative are
\begin{align}\label{zetauk}
	u_k (\tau) = \frac{iH}{\sqrt{4\epsilon c_s k^3}} (1+i c_s k \tau) e^{-ic_s k \tau}~,
	\qquad
	u'_k(\tau) = \frac{iH}{\sqrt{4\epsilon c_s k^3}} c_s^2 k^2 \tau e^{-ic_s k\tau}~.
\end{align}
The two-point correlation function can be written as
\begin{align}
	\langle \zeta_{\mathbf k_1}\zeta_{\mathbf k_2} \rangle' = \frac{2\pi^2}{k^3} \Delta_\zeta^2,
	\qquad 
	\Delta_\zeta^2 = \frac{1}{8\pi^2} \frac{H^2}{c_s \epsilon}~.
\end{align}
 
\subsection{The rule to relate external mode functions}
At $\tau=0$, the difference between the mode function of $\zeta$ for inflation and the mode function of $\phi$ for Minkowski spacetime is just a constant. Using \eqref{flatuk} and \eqref{zetauk}, we have
\begin{align}
u_k(0)_{\rm inflation} = \frac{iH}{\sqrt{2 \epsilon c_s} k}\, u_k(0)_{\rm flat}\, .
\end{align}
	For example, for a three-point function with external momenta $k_1$, $k_2$ and $k_3$, to translate the Minkowski correlation function to that of inflation, for the external lines, we have to add a factor
\begin{align}
\bigg(\frac{i H}{\sqrt{2\epsilon c_s}}\bigg)^3 \frac{1}{k_1 k_2 k_3}~.
\end{align}

\subsection{The rule to relate Hamiltonians}
In order to compute the three-point function, we need to perturb the Hamiltonian to the third order. If we focus on the part with non-trivial sound speed, the third order interaction Hamiltonian is
\begin{align}
H_3 = H_{\zeta'^3} + H_{\zeta' (\partial \zeta)^2}	~,
\end{align}
where 
\begin{align}
H_{\zeta'^3} (\tau) & = \int d^3x\,2 a \frac{\lambda}{H^3}  \zeta'^3~, \\
H_{\zeta' (\partial \zeta)^2} (\tau) & = -  \int d^3x\,a \frac{\Sigma}{H^3} (1- c_s^2) \zeta' (\partial \zeta)^2~.
\end{align}
For simplicity, in what follows we will suppress the momentum-conserving delta functions in the Hamiltonians in momentum space.
\begin{itemize}
\item The Minkowski Hamiltonian \\
\begin{itemize}
\item 	For $H_{\phi^3}$, if all fields in it contract with the field on its left hand side, it corresponds to the following integral,
\begin{align}
H_{\phi^3} \rightarrow \int_{-\infty}^0d\tau\, \frac{\lambda_3}{2 \sqrt{2} \sqrt{k_1 k_2 k_3} } e^{i k_{123} \tau}
\end{align}
\end{itemize} 
	\item The inflationary Hamiltonian\\
\begin{itemize}
		\item For $H_{\zeta'^3}$, if all fields in it contract with the field on the left hand side, it corresponds to the following integral,
		\begin{align}
			H_{\zeta'^3}\rightarrow \int_{-\infty}^{0} -2 \frac{\lambda}{H} \frac{i c_s^{3/2} e^{i k_{123} \tau} \sqrt{k_1 k_2 k_3} \tau^2}{8 \epsilon^{3/2} } d\tau~.
		\end{align}
		Comparing with the integral expression of $H_{\phi^3}$, we have
		\begin{align}
			H_{\zeta'^3} \rightarrow \frac{i c_s k_1 k_2 k_3 \lambda}{\sqrt{2} H \epsilon^{3/2} \lambda_3 } \frac{\partial^2}{\partial k_{123}^2} H_{\phi^3}~.
		\end{align}
		Thus the corresponding relation between the inflationary and Minkowski three-point functions is
\begin{align}\label{eq:3pt3a}
\langle \zeta_{\mathbf k_1}\zeta_{\mathbf k_2}\zeta_{\mathbf k_3} \rangle_{\rm inflation} = \bigg(\frac{i H}{\sqrt{2 \epsilon c_s}}\bigg)^3 \frac{1}{k_1 k_2 k_3} \frac{i c_s k_1 k_2 k_3 \lambda}{\sqrt{2} H \epsilon^{3/2} \lambda_3 } \frac{\partial^2}{\partial k_{123}^2} \langle \phi_{\mathbf k_1}\phi_{\mathbf k_2}\phi_{\mathbf k_3} \rangle_{\rm flat}~.
\end{align} 
		If all fields in it contract with the field on the right hand side, we will get the complex conjugate instead
		\begin{align}
		H_{\zeta'^3} \rightarrow -\frac{i c_s k_1 k_2 k_3 \lambda}{\sqrt{2} H \epsilon^{3/2} \lambda_3 } \frac{\partial^2}{\partial k_{123}^2} H_{\phi^3}~.
		\end{align}  
		If we use it to compute the exchange diagram, since there are only two external legs attached to the vertex, we should make the substitution $k_{123}\rightarrow k_{12}$. If $k_1$ and $k_2$ contract with the field on the left hand side and $k_3$ contracts with the field on the right hand side, it gives the same operator. 
		\item $H_{\zeta' (\partial \zeta)^2}$, if all fields in it contract with the field on the left hand side, it corresponds to the following integral,
		\begin{align}
			H_{\zeta' (\partial \zeta)^2} (k_1,k_2,k_3,k_{123}) \rightarrow \int_{-\infty}^{0} \frac{i(-1+c_s^2)e^{i k_{123}\tau} \sqrt{k_1} \mathbf k_2\cdot \mathbf k_3 \Sigma (1-i k_2\tau) (1-i k_3\tau)  }{8 H k_2^{3/2} k_3^{3/2} \epsilon^{3/2} c_s^{1/2} } d\tau~.
		\end{align}
		Here $k_1$ corresponds to the momentum of $\dot\zeta$. $k_2$ and $k_3$ correspond to the momenta of the second and third $\partial \zeta$, respectively. Comparing with the integral expression of $H_{\phi^3}$, we have
		\begin{align}
			H_{\zeta' (\partial \zeta)^2}(k_1,k_2,k_3,k_{123}) \rightarrow \frac{i (-1+c_s^2) k_1 \mathbf k_2\cdot \mathbf k_3\Sigma}{2\sqrt{2} \sqrt{c_s} H k_2 k_3 \epsilon^{3/2} \lambda_3 } \bigg( 1- (k_2+k_3) \frac{\partial}{\partial k_{123}} + k_2 k_3 \frac{\partial^2}{\partial k_{123}^2} \bigg) H_{\phi^3}~.
		\end{align}
		Thus the corresponding relation between the inflationary and Minkowski three-point functions is
		\begin{align}\label{eq:3ptdzpzpz}
		\nonumber
		& \langle \zeta_{\mathbf k_1}\zeta_{\mathbf k_2}\zeta_{\mathbf k_3} \rangle_{\rm inflation} \\ \nonumber
		& = \bigg(\frac{i H}{\sqrt{2 \epsilon c_s}}\bigg)^3 \frac{1}{k_1 k_2 k_3} \frac{i (-1+c_s^2) k_1 \mathbf k_2\cdot \mathbf k_3\Sigma}{2\sqrt{2} \sqrt{c_s} H k_2 k_3 \epsilon^{3/2} \lambda_3 } \bigg( 1- (k_2+k_3) \frac{\partial}{\partial k_{123}} + k_2 k_3 \frac{\partial^2}{\partial k_{123}^2} \bigg) \\
		& \langle \phi_{\mathbf k_1}\phi_{\mathbf k_2}\phi_{\mathbf k_3} \rangle_{\rm flat} \times \frac{1}{6} +{5\,\,\rm permutations}~.
		\end{align}
The factor of $1/6$ and the five permutations correspond to that actually this Hamiltonian is not symmetric in $k_1$, $k_2$ and $k_3$. When we want to use it in the scalar diagram, we should substitute $k_1$ with the momentum associated with the $\zeta'$ mode function whereas $k_2$ and $k_3$ with the $\partial \zeta$ mode function. $k_{123}$ should be substituted with the sum over the momenta of the external legs of this Hamiltonian. 

The relation \eqref{eq:3ptdzpzpz} is the final result to relate the inflationary and Minkowski leading order three-point functions for the $\zeta' (\partial\zeta)^2$ interaction, because here all the fields in the interaction Hamiltonian contract with fields to the left. However, for beyond-leading order calculations, or higher-point correlation functions, we may encounter the possibility that some fields in the interaction Hamiltonian contracts to the left and some contracts to the right. For these situations, we will also need the rules below:

		\item $H_{\zeta' (\partial \zeta)^2}$, if all fields in it contract with the field on the right hand side, it corresponds to the following integral,
		\begin{align}
		H_{\zeta' (\partial \zeta)^2} (k_1,k_2,k_3,k_{123}) \rightarrow \int_{-\infty}^{0} \frac{-i(-1+c_s^2)e^{-i k_{123}\tau} \sqrt{k_1} \mathbf k_2\cdot \mathbf k_3 \Sigma (1+i k_2\tau) (1+i k_3\tau)  }{8 H k_2^{3/2} k_3^{3/2} \epsilon^{3/2} c_s^{1/2} } d\tau~.
		\end{align}
		Here $k_1$ corresponds to the momentum of $\dot\zeta$. $k_2$ and $k_3$ correspond to the momenta of the second and third $\partial \zeta$, respectively. Comparing with $H_{\phi^3}$, we have
		\begin{align}
		H_{\zeta' (\partial \zeta)^2}(k_1,k_2,k_3,k_{123}) \rightarrow \frac{-i (-1+c_s^2) k_1 \mathbf k_2\cdot \mathbf k_3\Sigma}{2\sqrt{2} \sqrt{c_s} H k_2 k_3 \epsilon^{3/2} \lambda_3 } \bigg( 1- (k_2+k_3) \frac{\partial}{\partial k_{123}} + k_2 k_3 \frac{\partial^2}{\partial k_{123}^2} \bigg) H_{\phi^3}~.
		\end{align}
		The usage of it is the same as $H_{\zeta' (\partial \zeta)^2}(k_1,k_2,k_3,k_{123})$. 
		\item $H_{\zeta' (\partial \zeta)^2}$, two of them contracts with the field on the left and one of them contracts with the field on the right.
		\begin{itemize}
			\item $\zeta'$ contracts with right hand side
			\begin{align}
				H_{\zeta' (\partial \zeta)^2} (k_1,k_2,k_I,k_{12}) \rightarrow \int_{-\infty}^{0} \frac{-i(-1+c_s^2)e^{i (k_{12}-k_I)\tau} \sqrt{k_I} \mathbf k_1\cdot \mathbf k_2 \Sigma (1-i k_1\tau) (1-i k_2\tau)  }{8 H k_1^{3/2} k_2^{3/2} \epsilon^{3/2} c_s^{1/2} } d\tau~.
			\end{align}
			Comparing with the integral expression of $H_{\phi^3}$, we have
			\begin{align}
			H_{\zeta' (\partial \zeta)^2}(k_1,k_2,k_I,k_{12}) \rightarrow \frac{-i (-1+c_s^2) k_I \mathbf k_1\cdot \mathbf k_2\Sigma}{2\sqrt{2} \sqrt{c_s} H k_1 k_2 \epsilon^{3/2} \lambda_3 } \bigg( 1- (k_1+k_2) \frac{\partial}{\partial k_{12}} + k_1 k_2 \frac{\partial^2}{\partial k_{12}^2} \bigg) H_{\phi^3}~.
			\end{align}
			\item $\partial\zeta$ contracts with right hand side
			\begin{align}
				H_{\zeta' (\partial \zeta)^2} (k_1,k_2,k_I,k_{12}) \rightarrow \int_{-\infty}^{0} \frac{-i(-1+c_s^2)e^{i (k_{12}-k_I) \tau} \sqrt{k_1} \mathbf k_2\cdot \mathbf k_I \Sigma (1-i k_2\tau) (1-i k_I\tau)  }{8 H k_2^{3/2} k_I^{3/2} \epsilon^{3/2} c_s^{1/2} } d\tau~.
			\end{align}
			Comparing with the integral expression of $H_{\phi^3}$, we have
			\begin{align}
			H_{\zeta' (\partial \zeta)^2}(k_1,k_2,k_I,k_{12}) \rightarrow \frac{-i (-1+c_s^2) k_1 \mathbf k_2\cdot \mathbf k_I\Sigma}{2\sqrt{2} \sqrt{c_s} H k_2 k_I \epsilon^{3/2} \lambda_3 } \bigg( 1- (k_2+k_I) \frac{\partial}{\partial k_{12}} + k_2 k_I \frac{\partial^2}{\partial k_{12}^2} \bigg) H_{\phi^3}~.
			\end{align}
		\end{itemize}
	\end{itemize}
	\end{itemize} 
With these rules, one can get the relations for the four-point function. Details are left to Appendix \ref{appendA}. In general, this type of method is applicable for any $n$-point correlation functions at the tree level.

For loop diagrams, we expect similar symmetry-breaking operators for the integrand. However, the symmetry-breaking operators are expected to depend on the free momentum running in the loop. Thus we cannot extract the symmetry-breaking operator out from the loop integral with the current technique.

Note that if we would like to study some parts of a Feynman diagram (for example, studying the polology of a propagator) instead of a whole diagram, we may choose to apply the symmetry-breaking operators to a selection of relevant vertices only.

Also, we present the extension of this method for bouncing cosmology in Appendix \ref{applicationtobounce}.

\section{A Recursion Relation}\label{recursionrelation} 
In~\cite{Arkani-Hamed:2017fdk}, various recursion relations of conformal scalars are derived for the wave function of the de Sitter universe~\cite{Hartle:1983ai,Hertog:2011ky,Bunch:1978yq,Chernikov:1968zm}. Here we first use their method to obtain a recursion relation for massless fields in flat space thanks to the similarity between these two cases. Then, we apply symmetry breaking operators to obtain the corresponding recursion relation for massless fields for inflation.

For Minkowski space with an artificial ``future boundary'' at $\tau\rightarrow 0$, the wave function of the universe takes the form
\begin{align}\nonumber
	\Psi = {\rm exp} \bigg[ & \frac{1}{2!} \int d^3 z_1 \int d^3 z_2 \phi (z_1) \phi(z_2) \hat \psi_2 (z) \\ \nonumber
	 + & \frac{1}{3!} \int d^3 z_1 \int d^3 z_2 \int d^3 z_3 \phi (z_1) \phi(z_2) \phi (z_3) \hat \psi_3 (z) \\ 
	 + & \frac{1}{4!} \int d^3 z_1 \int d^3 z_2 \int d^3 z_3 \int d^3 z_4 \phi (z_1) \phi(z_2) \phi(z_3) \phi(z_4) \hat \psi_4 (z)  \bigg].
\end{align}
In the following, we work in the momentum space and $\hat\psi_n$ denotes the corresponding quantity in the momentum space with momentum conservation delta function striped. The $n$-point correlation functions at a fixed time slice $\tau_c$ can be obtained once we know the form of the wave function in the following way
\begin{align}
	\langle \phi_{\mathbf k_1} \cdots \phi_{\mathbf k_n} \rangle = \frac{\int\prod_{\mathbf k} d\phi_{\mathbf k} |\Psi[\phi_{\mathbf k},\tau_c]|^2 \phi_{\mathbf k_1}\cdots\phi_{\mathbf k_n}}{\int\prod_{\mathbf k} d\phi_{\mathbf k} |\Psi[\phi_{\mathbf k},\tau_c]|^2}
\end{align}
Using the Gaussian integral, the following dictionary can be established between the inflationary correlation function and the wave function of the universe,
\begin{align}
	\langle \phi_{\mathbf k}  \phi_{-\mathbf k} \rangle' = \frac{1}{-2{\rm Re} \hat \psi_2 }~.
\end{align}
The three-point function can be calculated as
\begin{align}
	\langle \phi_{\mathbf k_1} \phi_{\mathbf k_2} \phi_{\mathbf k_3} \rangle' = \frac{2{\rm Re} \hat \psi_3}{ (-2{\rm Re} \hat \psi_2)(-2{\rm Re} \hat \psi_2)(-2{\rm Re} \hat \psi_2) }~.
\end{align}
The four-point function with a scalar exchange of $\mathbf k_I$ can be calculated as
\begin{align}\nonumber
& \langle \phi_{\mathbf k_1} \phi_{\mathbf k_2} \phi_{\mathbf k_3} \phi_{\mathbf k_4} \rangle' \\ \label{4ptwavefunctionvscorrelation}
& = \frac{2{\rm Re} \hat \psi_4 }{(-2{\rm Re} \hat \psi_2) (-2{\rm Re} \hat \psi_2)(-2{\rm Re} \hat \psi_2)(-2{\rm Re} \hat \psi_2) } + \frac{2{\rm Re} \hat \psi_3 2{\rm Re} \hat \psi_3}{ (-2{\rm Re} \hat \psi_2)(-2{\rm Re} \hat \psi_2)(-2{\rm Re} \hat \psi_2)(-2{\rm Re} \hat \psi_2)(-2{\rm Re} \hat \psi_2) }~.
\end{align}  
$\hat \psi_4$ is contributed by the scalar exchange diagram and be calculated as 
\begin{align}
\hat \psi_4 = \int_{-\infty}^0 d\tau_1 \int_{-\infty}^0 d\tau_2 H_{k_1}(\tau_1)H_{k_2}(\tau_1)H_{k_3}(\tau_2)H_{k_4}(\tau_2)G_{k_I}(\tau_1,\tau_2)~,
\end{align}
where $H_k(\tau)$ and $G_k(\tau_1,\tau_2)$ are bulk-boundary and bulk-bulk propagators defined as
\begin{align}
	H_k (\tau) & = e^{i k \tau} \\
	G_k (\tau_1,\tau_2) & = \frac{1}{2 k} [e^{-i k (\tau_1-\tau_2)}\Theta(\tau_1-\tau_2) + e^{i k(\tau_1-\tau_2)}\Theta(\tau_2-\tau_1) - e^{i k (\tau_1+\tau_2)}  ]~,
\end{align}
where the purpose of the last term in the bulk-bulk propagator is to force its value vanish on the boundary.
The $\hat \psi_4$ can be calculated in the following way
\begin{align}
\hat \psi_4 = \int_{-\infty}^0 d\tau_1 \int_{-\infty}^0 d\tau_2 e^{i k_{34}\tau_2}  e^{i k_{12}\tau_1}  \frac{1}{2k_I} \bigg[ e^{-i k_I (\tau_1-\tau_2)} \Theta(\tau_1-\tau_2) + e^{i k_I(\tau_1-\tau_2)} \Theta(\tau_2-\tau_1) -e^{i k_I (\tau_1+\tau_2)} \bigg] ~.
\end{align}
If we integrate $\tau_1$ first, we have
\begin{align}\label{masteridentity}
\hat \psi_4  = \frac{1}{k_{12}^2- k_I^2} \bigg(\int_{-\infty}^{0} d\tau_2 \frac{1}{i} e^{i(k_{12}+k_{34})\tau_2} - \int_{-\infty}^{0} d\tau_2 \frac{1}{i} e^{i(k_I+k_{34})\tau_2} \bigg)~.
\end{align} 
In~\cite{Arkani-Hamed:2017fdk}, it is noted that the right hand side is just the difference of two three-point vertices of the wave function $\hat\psi_3$, with shifted momenta. And this observation can be applied recursively to obtain a relation between the $n$-point vertex of the wave function and the difference of many three-point vertexes with shifted momenta. Thus, one can regard $\hat\psi_3$ as a fundamental building block of the high-point vertices of the wave function. This is in analogous to the fact that in CFT, the three-point function is the fundamental building block; in AdS/CFT, the cubic Witten diagram (triple-K integral~\cite{Bzowski:2013sza,Bzowski:2015pba,Bzowski:2015yxv}) is the fundamental building block; and in the scattering amplitude literature, the MHV amplitude is the building block. Thus the significance of finding an inflationary correlation counter part of this kind of recursion relation not only lies in the interest of simplifying calculations, but also in theoretically understanding the mathematical structure of the correlation functions.

Making use of \eqref{4ptwavefunctionvscorrelation}, Equation \eqref{masteridentity} can be written as
\begin{align}\nonumber
& \frac{ \langle \phi_{\mathbf k_1}\phi_{\mathbf k_2}\phi_{\mathbf k_3}\phi_{\mathbf k_4}  \rangle'}{2 \langle \phi_{\mathbf k_1}\phi_{-\mathbf k_1} \rangle'\langle \phi_{\mathbf k_2}\phi_{-\mathbf k_2} \rangle'\langle \phi_{\mathbf k_3}\phi_{-\mathbf k_3} \rangle'\langle \phi_{\mathbf k_4}\phi_{-\mathbf k_4} \rangle'} =  \frac{ \langle \phi_{\mathbf k_1}\phi_{\mathbf k_2}\phi_{\mathbf k_I}\rangle'\langle\phi_{-\mathbf k_I}\phi_{\mathbf k_3}\phi_{\mathbf k_4}  \rangle' }{2\langle \phi_{\mathbf k_1}\phi_{-\mathbf k_1} \rangle'\langle \phi_{\mathbf k_2}\phi_{-\mathbf k_2} \rangle'\langle \phi_{\mathbf k_3}\phi_{-\mathbf k_3} \rangle'\langle \phi_{\mathbf k_4}\phi_{-\mathbf k_4} \rangle'\langle \phi_{\mathbf k_I}\phi_{-\mathbf k_I} \rangle'}  \\ \nonumber
&  + \frac{1}{k_I^2 - k_{12}^2} \bigg[ \frac{-\langle \phi_{\mathbf k_1}\phi_{\mathbf k_3}\phi_{\mathbf k_4} \rangle'}{2\langle \phi_{\mathbf k_1}\phi_{-\mathbf k_1}\rangle'  \langle \phi_{\mathbf k_3}\phi_{-\mathbf k_3} \rangle'\langle \phi_{\mathbf k_4}\phi_{-\mathbf k_4} \rangle'} \bigg|_{\lambda_3\rightarrow \lambda_3^2, k_1\rightarrow k_{12}} - \frac{-\langle \phi_{\mathbf k_3} \phi_{\mathbf k_4} \phi_{\mathbf k_I} \rangle'}{2\langle \phi_{\mathbf k_3} \phi_{-\mathbf k_3} \rangle'\langle \phi_{\mathbf k_4} \phi_{-\mathbf k_4} \rangle'\langle \phi_{\mathbf k_I} \phi_{-\mathbf k_I} \rangle'}\bigg|_{\lambda_3\rightarrow \lambda_3^2 } \bigg]\\
& + {\rm permutations}~.
\end{align}
Let's take $H_{\zeta'^3}$ interaction as an example, acting $\frac{H^2 k_I^2 \lambda^2}{8 \epsilon^5 \lambda_3^2}\frac{\partial^2}{\partial k_{12}^2}\frac{\partial^2}{\partial k_{34}^2}$ on both sides of the correlation, and after some simplification, we have\footnote{Here the direction of $\mathbf{k}_{12}$ is not defined or actually used. }
\begin{align}\nonumber
& \langle \zeta_{\mathbf k_1}\zeta_{\mathbf k_2}\zeta_{\mathbf k_3}\zeta_{\mathbf k_4}  \rangle'_{\rm inflation} = \frac{  \langle \zeta_{\mathbf k_1}\zeta_{\mathbf k_2}\zeta_{\mathbf k_I}\rangle'_{\rm inflation}\langle\zeta_{-\mathbf k_I}\zeta_{\mathbf k_3}\zeta_{\mathbf k_4}  \rangle'_{\rm inflation} }{ \langle \zeta_{\mathbf k_I}\zeta_{-\mathbf k_I} \rangle'_{\rm inflation} } \\ \nonumber
& + \bigg(  \frac{k_I^2\lambda c_s^{3/2}}{\epsilon\lambda_3^2 H^2}  \frac{\partial^2}{\partial k_{12}^2} \bigg\{ \frac{1}{k_I^2-k_{12}^2} \frac{-\langle\zeta_{\mathbf k_{12}} \zeta_{\mathbf k_3} \zeta_{\mathbf k_4} \rangle'_{\rm inflation}}{\langle \zeta_{\mathbf k_{12}}\zeta_{-\mathbf k_{12}} \rangle'_{\rm inflation} }  \frac{1}{k_{12}^2}  \bigg\}  - H \frac{\lambda c_s^{3/2}}{\epsilon \lambda_3^2 H^2} \frac{\partial^2}{\partial_{k_{12}}^2} \bigg\{ \frac{1}{k_I^2-k_{12}^2} \frac{-\langle \zeta_{\mathbf k_3} \zeta_{\mathbf k_4} \zeta_{\mathbf k_I}\rangle'_{\rm inflation}}{ \langle \zeta_{\mathbf k_I} \zeta_{-\mathbf k_I} \rangle'_{\rm inflation}} \bigg\} \bigg) \\  \label{recursionrelation2}
&  \langle \zeta_{\mathbf k_1}\zeta_{-\mathbf k_1} \rangle'_{\rm inflation} \langle \zeta_{\mathbf k_2}\zeta_{-\mathbf k_2} \rangle'_{\rm inflation}   {k_1^2 k_2^2} + {\rm permutations} ~.
\end{align}
Here $\langle \zeta_{\mathbf k_{12}}\zeta_{\mathbf k_3}\zeta_{\mathbf k_4} \rangle'$ is not a physical correlation function. This is because of the lack of momentum conservation: $\mathbf{k}_{12}+\mathbf{k}_3+\mathbf{k}_4\neq 0$. Nevertheless, for the particular $\dot\zeta^3$ interaction (and in general interactions without dot products of spatial derivatives), the three-point function does not depend on the direction of its momenta. Thus $\langle \zeta_{\mathbf k_{12}}\zeta_{\mathbf k_3}\zeta_{\mathbf k_4} \rangle' = \langle \zeta_{ k_{12}}\zeta_{ k_3}\zeta_{ k_4} \rangle'$. As a result, to test this recursion relation, one can search for correlation with momenta of size $k_{12}$, $k_3$ and $k_4$ in the sky as long as they satisfy the triangular inequalities.

On the other hand, one can also restrict the momenta configuration to the collinear limit\footnote{In cosmology, the collinear relations (which reduce to folded relations when considering the three-point functions) is known to appear in the context of non-Bunch-Davies vacua~\cite{Chen:2006nt, Chen:2009bc} and its decay~\cite{Jiang:2015hfa, Jiang:2016nok}. This is because the decay of non-Bunch-Davies modes and the resulting cosmological correlations are dominated by early times deep inside the Hubble radius, where the metric can be approximated by the Minkowski metric implicitly. Here, we are considering Bunch-Davies initial condition thus we are getting a different observation in the collinear limit.} where two external momenta (say, $\mathbf{k}_1$ and $\mathbf{k}_2$) are in the same direction. In this limit, $\mathbf{k}_1+\mathbf{k}_2 \rightarrow \mathbf{k}_{12}$ and thus the vectors $\mathbf k_{12}$, $\mathbf k_3$, $\mathbf k_4$ satisfy momentum conservation. In this limit, the factor $k_I^2-k_{12}^2$ in the denominator blows up. Thus we have to Taylor-expand the denominator and the numerator (similar to the L'Hospital's rule, but the existence of the operator $\partial_{k_{12}}^2$ forces us to expand to third order in the Taylor series to get a meaningful result). The recursion relation then becomes
\begin{align}\nonumber
& \langle \zeta_{\mathbf k_1}\zeta_{\mathbf k_2}\zeta_{\mathbf k_3}\zeta_{\mathbf k_4}  \rangle'_{\rm inflation} =  \frac{  \langle \zeta_{\mathbf k_1}\zeta_{\mathbf k_2}\zeta_{\mathbf k_I}\rangle'_{\rm inflation}\langle\zeta_{-\mathbf k_I}\zeta_{\mathbf k_3}\zeta_{\mathbf k_4}  \rangle'_{\rm inflation} }{ \langle \zeta_{\mathbf k_I}\zeta_{-\mathbf k_I} \rangle'_{\rm inflation} } \\ \nonumber 
&+ \frac{\lambda c_s^{3/2}}{\epsilon\lambda_3^2}  \bigg\{ \frac{1}{-4 k_I} \bigg[ (-  k_I) f^{(2)}_{\rm  inflation} +   f^{(1)}_{\rm  inflation} + \frac{2}{3}  k_I^2 f^{(3)}_{\rm  inflation} \bigg] \bigg\} + {\rm permutations} ~,
\end{align}
where
\begin{align}
f^{(n)}_{\rm inflation}  = \frac{\partial}{\partial\varepsilon^n} \bigg[ \frac{-\langle \zeta_{\mathbf k_{12}}\zeta_{\mathbf k_3}\zeta_{\mathbf k_4} \rangle'_{\rm inflation}}{ \langle \zeta_{\mathbf k_{12}}\zeta_{-\mathbf k_{12}}\rangle'_{\rm inflation} \frac{k_{12}^2}{H^2} } \bigg|_{k_{12}\rightarrow k_I+\varepsilon} \bigg] \langle \zeta_{\mathbf k_1}\zeta_{-\mathbf k_1} \rangle'_{\rm inflation} \langle \zeta_{\mathbf k_2}\zeta_{-\mathbf k_2} \rangle'_{\rm inflation} \frac{k_{1}^2}{H^2}\frac{k_{2}^2}{H^2}~.
\end{align}
Note that the curvature perturbation is massless. Thus in flat space, the collinear limit corresponds to the situation where an internal propagator is on-shell if the external lines are on-shell. In the context of cosmology, the on-shell condition is not obvious because we are forced to consider space and time differently. Also we are considering correlation functions without making them into scattering amplitudes. Nevertheless, our method of extracting the symmetry-breaking operator allows us to find out such a collinear relation. It remains interesting to see if the result can be further connected to recent studies of scattering amplitudes (see for example~\cite{Nandan:2016ohb} and references therein for the related flat space amplitudes).

Although we have been working in a particular model, the recursion relation has a range of generality:
\begin{itemize}
  \item The relation is only sensitive to the vertex connecting $k_1$ and $k_2$ to internal propagators. No information on the interaction structure of the rest part of the Feynman diagram is needed. Thus we can have similar relations from a complicated $n$-point correlation function with general interactions, as long as the vertex attached to $k_1$ and $k_2$ has the same interaction structure as above.
  \item Here we have considered the interaction $H_{\zeta'^3}$ for the corresponding vertex. If a different Hamiltonian is used, a similar relation may apply, where different kinds of operators are applied onto the correlation functions.
\end{itemize}
We plan to leave a detailed study of this recursion relation, its generalizations and cosmological implications to a future work.
	
\section{Conclusion and Outlook}\label{conclusionandoutlook} 

We have shown that massless cosmological correlation functions can be obtained by acting symmetry breaking operators to Minkowski correlation functions. We have derived the operators corresponding to general single field inflation as an example. Once written in terms of the Minkowski correlation function,  properties in Minkowski spacetime can be translated to relations in cosmology. 

There are many interesting possibilities to explore. Here we list a few examples. We hope to address some of these possibilities in the future.

\begin{itemize}
  \item We have not taken advantages of Minkowski symmetries. It would be helpful to write the targeting Minkowski correlation functions in obviously 4-dimensional covariant formalism. More structures on the cosmological correlation functions may be uncovered due to the obvious symmetries.
  \item We have focused on massless scalar fields. It is interesting to generalize the method to include massless fields with higher spins, especially primordial gravitational waves. Also, acting an operator on a scalar exchange diagram can give us the result of a high spin massless field exchange diagram, which simplifies the previous result of~\cite{Seery:2008ax}. Similar technique can also be applied to AdS/CFT to simplify the calculation about Witten diagram~\cite{Hiroshi}.
  \item We have focused on correlation functions for Minkowski and cosmological types. However, in flat space, much more relations are known for scattering amplitudes compared to general correlation functions. Conceptually, we expect that similar techniques should apply for cosmology. This is because on super-Hubble scales, the fluctuations get frozen and become classical. The classical fluctuations should be considered on-shell as their quantum components become irrelevant. It is thus valuable to construct a LSZ-type formalism to pick up the on-shell component of the correlation function and take advantage of the on-shell condition for further analysize the structure of the cosmological correlation functions, for example, unitarity, analyticity, causality~\cite{Adams:2006sv,Baumann:2015nta} and so on. 
  \item Our current method only applies for massless fields (in fact also conformal fields but that is almost trivial). It is important to seek for alternative ways to relate cosmology with the Minkowski correlation function, with the generality to include fields with arbitrary mass. One hope is by taking the $k_t\rightarrow 0$ limit~\cite{Arkani-Hamed:2015bza, Arkani-Hamed:2017fdk} in the analytically continued momentum space. The reason is that in the very early universe deep inside the Hubble radius, the spacetime is approximately Minkowski. In AdS/CFT, this is famously known as a bulk point singularity~\cite{Raju:2012zr}. It is interesting to see how far one can proceed in this direction and how much cosmological information can get recovered from this approach.
  \item We have relied on perturbation theory in searching for the symmetry breaking operators. It is interesting to seek for a non-perturbative description which is not organized in diagram-by-diagram basis. Such non-perturbative operators, if identified, may help in understanding the nonlinear physics of the large scale structure based on enhanced symmetries.
  \item It remains interesting to find a geometric implication of the known relations in cosmology. The polytope structure~\cite{Arkani-Hamed:2017fdk} offers us this possibility. Especially the de Sitter dilatation symmetry and special conformal symmetry can be visualized as some transformation on the cosmological polytope. A more ambitious goal is to find an object that automatically taking into account all the diagrams in the perturbation series like what is discovered about the amplituhedron~\cite{Arkani-Hamed:2013jha,Arkani-Hamed:2013kca,Arkani-Hamed:2017vfh} in Minkowski spacetime.
\end{itemize}

\section*{Acknowledgments} 
We thank Dionysios Anninos, Andrew Cohen, Hayden Lee, Shing Yan Li, Yubin Li and Toshifumi Noumi for useful discussions. This work is supported in part by ECS Grant 26300316 and GRF Grant 16301917 from the Research Grants Council of Hong Kong. XT is supported by the Qian San-Qiang Class in the University of Science and Technology of China.  SZ is supported by the Hong Kong PhD Fellowship Scheme (HKPFS) issued by the Research Grants Council (RGC) of Hong Kong.

\appendix
\section{Four-Point Function}\label{appendA}
In this appendix, we calculate the four-point function of general single field inflation which was originally obtained in~\cite{Chen:2009bc,Arroja:2009pd}. We used the symmetry-breaking operator formalism and show the simplifications therein.

For the four-point function, we have a factor contributed by the external legs, 
\begin{align}
\bigg(\frac{i H}{\sqrt{2\epsilon c_s}}\bigg)^4 \frac{1}{k_1 k_2 k_3 k_4}~.
\end{align} 
Also we need the following form of the Hamiltonian for $H_{\phi^4}$. If all field in it contract with the field on its left hand side, it corresponds to the following integral,
\begin{align}
	H_{\phi^4} \rightarrow \int_{-\infty}^0 \frac{\lambda_4}{4 \sqrt{k_1 k_2 k_3 k_4} } e^{i k_{1234} \tau} d\tau~.
\end{align}
\subsection{Contact-interaction Diagram}
We compute the four-point correlation function contributed by the contact-interaction diagram. This type of diagram was originally calculated in~\cite{Huang:2006eha,Arroja:2008ga}. We define the following parameter
\begin{align}
\mu \equiv \frac{1}{2} X^2 P_{,XX} + 2 X^3 P_{,XXX} + \frac{2}{3} X^4 P_{,XXXX}~.
\end{align}
The forth order Hamiltonian takes the form
\begin{align}
H_{4}   = H_{\zeta'^4} + H_{(\partial \zeta)^2 \zeta'^2} + H_{(\partial\zeta)^4 }~,
\end{align} 
where
\begin{align}
H_{\zeta'^4} (\tau) & = \frac{1}{H^4} \bigg(-\mu + 9 \frac{\lambda^2}{\Sigma}\bigg) \zeta'^4~, \\
H_{(\partial \zeta)^2 \zeta'^2} (\tau) & = \frac{1}{H^4} (3\lambda c_s^2 - \Sigma(1-c_s^2)) (\partial \zeta)^2 \zeta'^2~, \\
H_{(\partial\zeta)^4 } (\tau) & = \frac{1}{4H^4} \Sigma (- c_s^2 + c_s^4) (\partial\zeta)^4 ~.
\end{align} 
Now we derive the rules relating the inflationary Hamiltonian and the Minkowski Hamiltonian. For the inflationary correlation functions contributed by different interactions, we all use $\langle\cdots \rangle'_{\rm inflation}$ to avoid complexity of notation. The different contractions are classified in the following cases:
\begin{itemize} 
	\item $H_{\zeta'^4}$, if all fields in it contract with the field on the left hand side, it corresponds to the following integral,
	\begin{align}
	H_{\zeta'^4} \rightarrow \int_{-\infty}^{0} - \frac{c_s e^{i k_{1234} \tau } \sqrt{k_1 k_2 k_3 k_4} (-9\lambda^2+\mu\Sigma) \tau^4 }{16 \epsilon^2 \Sigma} d\tau ~.
	\end{align}
	Comparing with the integral expression of $H_{\phi^4}$, we have
	\begin{align}
	H_{\zeta'^4} \rightarrow - \frac{c_s k_1 k_2 k_3 k_4(-9\lambda^2+\mu\Sigma)}{4 \epsilon^2 \lambda \Sigma} \frac{\partial^4}{\partial k_{1234}^4} H_{\phi^4}~.
	\end{align}
	Thus the corresponding relation between the inflationary and Minkowski four-point functions is
	\begin{align}\label{eq:4ptdz4}
	\nonumber
	& \langle \zeta_{\mathbf k_1}\zeta_{\mathbf k_2}\zeta_{\mathbf k_3}\zeta_{\mathbf k_4} \rangle_{\rm inflation} \\
	& = \bigg(\frac{i H}{\sqrt{2\epsilon c_s}}\bigg)^4 \frac{1}{k_1 k_2 k_3 k_4} \bigg(- \frac{c_s k_1 k_2 k_3 k_4(-9\lambda^2+\mu\Sigma)}{4 \epsilon^2 \lambda \Sigma} \frac{\partial^4}{\partial k_{1234}^4}\bigg) \langle \phi_{\mathbf k_1}\phi_{\mathbf k_2}\phi_{\mathbf k_3}\phi_{\mathbf k_4} \rangle_{\rm flat}~.
	\end{align} 
	\item $H_{(\partial \zeta)^2 \zeta'^2}$, if all fields in it contract with the field on the left hand side, it corresponds to the following integral,
	\begin{align}
	H_{(\partial \zeta)^2 \zeta'^2} (k_1,k_2,k_3,k_4) \rightarrow \int_{-\infty}^{0} \frac{e^{i k_{1234}\tau} \sqrt{k_1} \sqrt{k_2} \mathbf k_3\cdot \mathbf k_4 (-\Sigma+c_s^2(3\lambda+\Sigma)) }{16 c_s k_3^{3/2} k_4^{3/2} \epsilon^2 } (\tau_1^2(-1) (1-i k_3\tau)(1-i k_4\tau) ) d\tau~.
	\end{align}
	Comparing with the integral expression of $H_{\phi^4}$, we have
	\begin{align}
	H_{(\partial \zeta)^2 \zeta'^2} (k_1,k_2,k_3,k_4) \rightarrow \frac{k_1 k_2 \mathbf k_3\cdot \mathbf k_4 (-\Sigma+c_s^2(3\lambda+\Sigma)) }{4 c_s k_3 k_4 \epsilon^2 \lambda_4} \frac{\partial^2}{\partial k_{1234}^2} \bigg(1-(k_3+k_4)\frac{\partial}{\partial k_{1234}} + k_3 k_4 \frac{\partial^2}{\partial k_{1234}^2} \bigg) H_{\phi^4} ~.
	\end{align}
	Thus the corresponding relation between the inflationary and Minkowski four-point functions is
	\begin{align}\label{eq:4ptpz2dz2}
	\nonumber
	& \langle \zeta_{\mathbf k_1}\zeta_{\mathbf k_2}\zeta_{\mathbf k_3}\zeta_{\mathbf k_4} \rangle_{\rm inflation} \\ \nonumber
	= &\bigg(\frac{i H}{\sqrt{2\epsilon c_s}}\bigg)^4 \frac{1}{k_1 k_2 k_3 k_4}  \frac{k_1 k_2 \mathbf k_3\cdot \mathbf k_4 (-\Sigma+c_s^2(3\lambda+\Sigma)) }{4 c_s k_3 k_4 \epsilon^2 \lambda_4} \\
	& \times\frac{\partial^2}{\partial k_{1234}^2}  \bigg(1-(k_3+k_4)\frac{\partial}{\partial k_{1234}} + k_3 k_4 \frac{\partial^2}{\partial k_{1234}^2} \bigg) \langle \phi_{\mathbf k_1}\phi_{\mathbf k_2}\phi_{\mathbf k_3}\phi_{\mathbf k_4} \rangle_{\rm flat} \times\frac{1}{24} +{\rm 23\,\, permutations}~.
	\end{align}
	\item $H_{(\partial\zeta)^4}$, if all fields in it contract with the field on the left hand side, it corresponds to the following integral,
	\begin{align}
	H_{(\partial\zeta)^4} \rightarrow \int_{-\infty}^{0} (-1) \frac{(-1+c_s^2) e^{i k_{1234} \tau} \mathbf k_1\cdot\mathbf k_2 \mathbf k_3\cdot \mathbf k_4 (1-i k_1\tau)(1-i k_2\tau)(1-i k_3\tau)(1-i k_4\tau) }{64 k_1^{3/2}k_2^{3/2}k_3^{3/2}k_4^{3/2} \epsilon^2 } d\tau ~.
	\end{align}
	Comparing with the integral expression of $H_{\phi^4}$, we have
	\begin{align}\nonumber
	H_{(\partial\zeta)^4}  \rightarrow &-\frac{(-1+c_s^2)\mathbf k_1\cdot \mathbf k_2 \mathbf k_3\cdot \mathbf k_4 \Sigma}{16 k_1 k_2 k_3 k_4 \epsilon \lambda_4}\\ 
	&\times \bigg[ 1 - (k_1+k_2+k_3+k_4) \frac{\partial}{\partial k_{1234}} + \sum_{i\neq j} k_i k_j \frac{\partial^2}{\partial k_{1234}^2}  - \sum_{i\neq j\neq k} k_i k_j k_k \frac{\partial^3}{\partial k_{123}^3} + k_1 k_2 k_3 k_4 \frac{\partial^4}{\partial k_{1234}^4} \bigg] H_{\phi^4}~.
	\end{align}
	Thus the corresponding relation between the inflationary and Minkowski four-point functions is
	\begin{align}\label{eq:4ptpz4}
	\nonumber
	& \langle \zeta_{\mathbf k_1}\zeta_{\mathbf k_2}\zeta_{\mathbf k_3}\zeta_{\mathbf k_4} \rangle_{\rm inflation} \\ \nonumber
	= &\bigg(\frac{i H}{\sqrt{2\epsilon c_s}}\bigg)^4 \frac{1}{k_1 k_2 k_3 k_4}  \bigg(-\frac{(-1+c_s^2)(\mathbf k_1\cdot \mathbf k_2) (\mathbf k_3\cdot \mathbf k_4) \Sigma}{16 k_1 k_2 k_3 k_4 \epsilon \lambda_4}\bigg)\\\nonumber
	& \times \bigg[ 1 - k_{1234}\frac{\partial}{\partial k_{1234}} + \sum_{i\neq j} k_i k_j \frac{\partial^2}{\partial k_{1234}^2}  - \sum_{i\neq j\neq k} k_i k_j k_k \frac{\partial^3}{\partial k_{123}^3} + k_1 k_2 k_3 k_4 \frac{\partial^4}{\partial k_{1234}^4} \bigg]  
	\\ &
	\times \langle \phi_{\mathbf k_1}\phi_{\mathbf k_2}\phi_{\mathbf k_3}\phi_{\mathbf k_4} \rangle_{\rm flat}
	\times\frac{1}{24} +{\rm 23\,\, permutations}~.
	\end{align}
\end{itemize}

\subsection{Scalar Exchange Diagram}
We evaluate the relation between inflationary four-point function contributed by the scalar exchange diagram and that of the Minkowski spacetime using the tools we already obtained in Section~\ref{applicationtoinflation}. 
\begin{itemize}
	\item Contribution from $H_{\zeta'^3}$ and $H_{\zeta'^3}$ 
	\begin{align}\nonumber
	\langle \zeta_{\mathbf k_1}\zeta_{\mathbf k_2}\zeta_{\mathbf k_3}\zeta_{\mathbf k_4} \rangle_{\rm inflation} & =   \int_{\tau_0}^{\tau} d\tau_1 \int_{\tau_0}^{\tau} d\tau_2 \langle 0 | H_{\zeta'^3} (\tau_1) \zeta_{\mathbf k_1}\zeta_{\mathbf k_2}\zeta_{\mathbf k_3}\zeta_{\mathbf k_4} H_{\zeta'^3} (\tau_2) | 0 \rangle \\
	& - 2 {\rm Re} \int_{\tau_0}^{\tau} d\tau_1 \int_{\tau_0}^{\tau_1} d\tau_2 \langle 0 |  \zeta_{\mathbf k_1}\zeta_{\mathbf k_2}\zeta_{\mathbf k_3}\zeta_{\mathbf k_4} H_{\zeta'^3} (\tau_1) H_{\zeta'^3} (\tau_2) | 0 \rangle~,
	\end{align}
	we take the initial time going to $-\infty$ and final time going to $0$.  
	The relation between the four-point function contributed by the $H_{\zeta'^3}$ and $H_{\zeta'^3}$ interaction and that of the Minkowski correlation function contributed by $H_{\phi^3}$ interaction is
	\begin{align}\nonumber
	\langle \zeta_{\mathbf k_1}\zeta_{\mathbf k_2}\zeta_{\mathbf k_3}\zeta_{\mathbf k_4} \rangle_{\rm inflation} & = \frac{c_s H^2 k_I^2 \lambda^2}{8\epsilon^5} 
	\frac{\partial^2}{\partial k_{12}^2} \frac{\partial^2}{\partial k_{34}^2} \langle \phi_{\mathbf k_1}\phi_{\mathbf k_2}\phi_{\mathbf k_3}\phi_{\mathbf k_4} \rangle'_{ {\rm flat noP}} \times 4  + {\rm permutations}~,
	\end{align}
	where the subscript ``noP" denotes the $\langle \phi_{\mathbf k_1}\phi_{\mathbf k_2}\phi_{\mathbf k_3}\phi_{\mathbf k_4} \rangle'_{\rm flat}$ without 23 permutations.

	\item Contribution from $H_{\zeta' (\partial \zeta)^2}$ and $H_{\zeta'^3}$  
	\begin{align}\nonumber
	\langle \zeta_{\mathbf k_1}\zeta_{\mathbf k_2}\zeta_{\mathbf k_3}\zeta_{\mathbf k_4} \rangle_{\rm inflation} & =   \int_{\tau_0}^{\tau} d\tau_1 \int_{\tau_0}^{\tau} d\tau_2 \langle 0 | H_{\zeta' (\partial \zeta)^2} (\tau_1) \zeta_{\mathbf k_1}\zeta_{\mathbf k_2}\zeta_{\mathbf k_3}\zeta_{\mathbf k_4} H_{\zeta'^3} (\tau_2) | 0 \rangle \\
	& - 2 {\rm Re} \int_{\tau_0}^{\tau} d\tau_1 \int_{\tau_0}^{\tau_1} d\tau_2 \langle 0 |  \zeta_{\mathbf k_1}\zeta_{\mathbf k_2}\zeta_{\mathbf k_3}\zeta_{\mathbf k_4} H_{\zeta' (\partial \zeta)^2} (\tau_1) H_{\zeta'^3} (\tau_2) | 0 \rangle~,
	\end{align}
	we take the initial time going to $-\infty$ and final time going to $0$. Then we have the relation between the four-point function contributed by the $H_{\zeta' (\partial \zeta)^2}$ and $H_{\zeta'^3}$ interaction and that of the Minkowski correlation function contributed by $H_{\phi^3}$ interaction
	\begin{itemize}
		\item $\zeta'$ contracts with $\zeta'$ 
		\begin{align}\nonumber
		\langle \zeta_{\mathbf k_1}\zeta_{\mathbf k_2}\zeta_{\mathbf k_3}\zeta_{\mathbf k_4} \rangle_{\rm inflation} & = \frac{(-1+c_s^2)H^2 \mathbf k_1\cdot \mathbf k_2 k_I^2 \lambda \Sigma}{3\times 16 k_1^2 k_2^2 c_s \epsilon^5} \frac{\partial^2}{\partial k_{34}^2} \bigg(1-k_{12}\frac{\partial}{\partial k_{12} } + k_1 k_2 \frac{\partial^2}{\partial k_{12}^2} \bigg) \\\nonumber
		& \langle \phi_{\mathbf k_1}\phi_{\mathbf k_2}\phi_{\mathbf k_3}\phi_{\mathbf k_4} \rangle'_{\rm flat{\rm noP}} \times 4 + {\rm permutations}~.
		\end{align}
		\item $\partial\zeta$ contracts with $\zeta'$ 
		\begin{itemize}
			\item Non-time-ordered part
			\begin{align}\nonumber
			\langle \zeta_{\mathbf k_1}\zeta_{\mathbf k_2}\zeta_{\mathbf k_3}\zeta_{\mathbf k_4} \rangle_{\rm inflation} & = \frac{2\times (-1+c_s^2)H^2 \mathbf k_I \cdot \mathbf k_2  \lambda \Sigma}{3\times 16   k_2^2 c_s \epsilon^5} \frac{\partial^2}{\partial k_{34}^2} \bigg(1-(k_{2}+k_I)\frac{\partial}{\partial k_{12} } + k_I k_2 \frac{\partial^2}{\partial k_{12}^2} \bigg) \\\nonumber
			& \langle \phi_{\mathbf k_1}\phi_{\mathbf k_2}\phi_{\mathbf k_3}\phi_{\mathbf k_4} \rangle'_{\rm flat{\rm noP}} \times 4 + {\rm permutations}~.
			\end{align}
			\item Time-ordered part
			\begin{align}\nonumber
			\langle \zeta_{\mathbf k_1}\zeta_{\mathbf k_2}\zeta_{\mathbf k_3}\zeta_{\mathbf k_4} \rangle_{\rm inflation} & = \frac{2\times (-1+c_s^2)H^2 \mathbf k_I \cdot \mathbf k_2  \lambda \Sigma}{3\times 16   k_2^2 c_s \epsilon^5} \frac{\partial^2}{\partial k_{34}^2} \bigg(1-(k_{2}-k_I)\frac{\partial}{\partial k_{12} } - k_I k_2 \frac{\partial^2}{\partial k_{12}^2} \bigg) \\\nonumber
			& \langle \phi_{\mathbf k_1}\phi_{\mathbf k_2}\phi_{\mathbf k_3}\phi_{\mathbf k_4} \rangle'_{\rm flat{\rm noP}} \times 4  + {\rm permutations}~.
			\end{align}
		\end{itemize}
	\end{itemize}
	\item Contribution from $H_{\zeta' (\partial \zeta)^2}$ and $H_{\zeta' (\partial \zeta)^2}$ 
	\begin{align}\nonumber
	\langle \zeta_{\mathbf k_1}\zeta_{\mathbf k_2}\zeta_{\mathbf k_3}\zeta_{\mathbf k_4} \rangle_{\rm inflation} & =   \int_{\tau_0}^{\tau} d\tau_1 \int_{\tau_0}^{\tau} d\tau_2 \langle 0 | H_{\zeta' (\partial \zeta} (\tau_1) \zeta_{\mathbf k_1}\zeta_{\mathbf k_2}\zeta_{\mathbf k_3}\zeta_{\mathbf k_4} H_{\zeta' (\partial \zeta} (\tau_2) | 0 \rangle \\
	& - 2 {\rm Re} \int_{\tau_0}^{\tau} d\tau_1 \int_{\tau_0}^{\tau_1} d\tau_2 \langle 0 |  \zeta_{\mathbf k_1}\zeta_{\mathbf k_2}\zeta_{\mathbf k_3}\zeta_{\mathbf k_4} H_{\zeta' (\partial \zeta} (\tau_1) H_{\zeta' (\partial \zeta} (\tau_2) | 0 \rangle~,
	\end{align}
	we take the initial time going to $-\infty$ and final time going to $0$. Then we have the relation between the four-point function contributed by the $H_{\zeta' (\partial \zeta)^2}$ and $H_{\zeta' (\partial \zeta)^2}$ interaction and that of the Minkowski correlation function contributed by $H_{\phi^3}$ interaction
	\begin{itemize}
		\item  $\zeta'$ contracts with $\zeta'$ \\
		\begin{align}\nonumber
		&\langle \zeta_{\mathbf k_1}\zeta_{\mathbf k_2}\zeta_{\mathbf k_3}\zeta_{\mathbf k_4} \rangle_{\rm inflation} \\\nonumber
		& = \frac{ (-1+c_s^2)^2 H^2 \mathbf k_1 \cdot \mathbf k_2 \mathbf k_3 \cdot \mathbf k_4  k_I^2 \Sigma^2 }{9\times 32   k_1^2 k_2^2 k_3^2 k_4^2 c_s^3 \epsilon^5} \bigg(1-(k_1+k_{2})\frac{\partial}{\partial k_{12} } + k_1 k_2 \frac{\partial^2}{\partial k_{12}^2} \bigg) \bigg(1-(k_3+k_{4})\frac{\partial}{\partial k_{34} } + k_3 k_4 \frac{\partial^2}{\partial k_{34}^2} \bigg) \\\nonumber
		& \langle \phi_{\mathbf k_1}\phi_{\mathbf k_2}\phi_{\mathbf k_3}\phi_{\mathbf k_4} \rangle'_{\rm flat{\rm noP}} \times 4 + {\rm permutations}~,
		\end{align}
		\item  $\zeta'$ contracts with $\partial\zeta$ \\ 
		\begin{align}\nonumber
		&\langle \zeta_{\mathbf k_1}\zeta_{\mathbf k_2}\zeta_{\mathbf k_3}\zeta_{\mathbf k_4} \rangle_{\rm inflation} \\ \nonumber
		& = \frac{ 2\times 2 (-1+c_s^2)^2 H^2 \mathbf k_1 \cdot \mathbf k_2 \mathbf k_I \cdot \mathbf k_4   \Sigma^2 }{9\times 32   k_1^2 k_2^2 k_4^2 c_s^3 \epsilon^5} \bigg(1-(k_1+k_{2})\frac{\partial}{\partial k_{12} } + k_1 k_2 \frac{\partial^2}{\partial k_{12}^2} \bigg) \bigg(1-(k_{4}+k_I)\frac{\partial}{\partial k_{34} } + k_I k_4 \frac{\partial^2}{\partial k_{34}^2} \bigg) \\\nonumber
		& \langle \phi_{\mathbf k_1}\phi_{\mathbf k_2}\phi_{\mathbf k_3}\phi_{\mathbf k_4} \rangle'_{\rm flat{\rm noP}} \times 4 + {\rm permutations}~,
		\end{align}
		\item  $\partial\zeta$ contracts with $\zeta'$\\ 
		\begin{itemize}
			\item Non-time-ordered part
			\begin{align}\nonumber
			&\langle \zeta_{\mathbf k_1}\zeta_{\mathbf k_2}\zeta_{\mathbf k_3}\zeta_{\mathbf k_4} \rangle_{\rm inflation} \\ \nonumber
			& = \frac{ 2\times 2 (-1+c_s^2)^2 H^2 \mathbf k_3 \cdot \mathbf k_4 \mathbf k_I \cdot \mathbf k_2   \Sigma^2 }{9\times 32   k_3^2 k_2^2 k_4^2 c_s^3 \epsilon^5} \bigg(1-(k_{2}+k_I)\frac{\partial}{\partial k_{12} } + k_I k_2 \frac{\partial^2}{\partial k_{12}^2} \bigg) \bigg(1-(k_3+k_{4})\frac{\partial}{\partial k_{34} } + k_3 k_4 \frac{\partial^2}{\partial k_{34}^2} \bigg) \\\nonumber
			& \langle \phi_{\mathbf k_1}\phi_{\mathbf k_2}\phi_{\mathbf k_3}\phi_{\mathbf k_4} \rangle'_{\rm flat{\rm noP}} \times 4 + {\rm permutations}~,
			\end{align}
			\item Time-ordered part
			\begin{align}\nonumber
			&\langle \zeta_{\mathbf k_1}\zeta_{\mathbf k_2}\zeta_{\mathbf k_3}\zeta_{\mathbf k_4} \rangle_{\rm inflation} \\ \nonumber
			& = \frac{ 2\times 2 (-1+c_s^2)^2 H^2 \mathbf k_3 \cdot \mathbf k_4 \mathbf k_I \cdot \mathbf k_2   \Sigma^2 }{9\times 32   k_3^2 k_2^2 k_4^2 c_s^3 \epsilon^5} \bigg(1-(k_{2}-k_I)\frac{\partial}{\partial k_{12} } - k_I k_2 \frac{\partial^2}{\partial k_{12}^2} \bigg) \bigg(1-(k_3+k_{4})\frac{\partial}{\partial k_{34} } + k_3 k_4 \frac{\partial^2}{\partial k_{34}^2} \bigg) \\\nonumber
			& \langle \phi_{\mathbf k_1}\phi_{\mathbf k_2}\phi_{\mathbf k_3}\phi_{\mathbf k_4} \rangle'_{\rm flat{\rm noP}} \times 4 + {\rm permutations}~,
			\end{align}
		\end{itemize}
		\item  $\partial\zeta$ contracts with $\partial\zeta$  
		\begin{itemize}
			\item Non-time-ordered part  
			\begin{align}\nonumber
			&\langle \zeta_{\mathbf k_1}\zeta_{\mathbf k_2}\zeta_{\mathbf k_3}\zeta_{\mathbf k_4} \rangle_{\rm inflation} = \frac{2\times 2 (-1+c_s^2)^2 \mathbf k_I \cdot \mathbf k_2 \mathbf k_I \cdot \mathbf k_4 \Sigma^2 }{9\times 32 k_2^2 k_4^2 k_I^2 \epsilon^5} \\\nonumber 
			& \bigg(1-(k_{2}+k_I)\frac{\partial}{\partial k_{12} } + k_I k_2 \frac{\partial^2}{\partial k_{12}^2} \bigg) \bigg(1-(k_{4}+k_I)\frac{\partial}{\partial k_{34} } + k_I k_4 \frac{\partial^2}{\partial k_{34}^2} \bigg) \\\nonumber
			& \langle \phi_{\mathbf k_1}\phi_{\mathbf k_2}\phi_{\mathbf k_3}\phi_{\mathbf k_4} \rangle'_{\rm flat{\rm noP}} \times 4 + {\rm permutations}~,
			\end{align} 
			\item Time-ordered part  
			\begin{align}\nonumber
			&\langle \zeta_{\mathbf k_1}\zeta_{\mathbf k_2}\zeta_{\mathbf k_3}\zeta_{\mathbf k_4} \rangle_{\rm inflation} = \frac{2\times 2 (-1+c_s^2)^2 \mathbf k_I \cdot \mathbf k_2 \mathbf k_I \cdot \mathbf k_4 \Sigma^2 }{9\times 32 k_2^2 k_4^2 k_I^2 \epsilon^5} \\\nonumber 
			& \bigg(1-(k_{2}-k_I)\frac{\partial}{\partial k_{12} } - k_I k_2 \frac{\partial^2}{\partial k_{12}^2} \bigg) \bigg(1-(k_{4}+k_I)\frac{\partial}{\partial k_{34} } + k_I k_4 \frac{\partial^2}{\partial k_{34}^2} \bigg) \\\nonumber
			& \langle \phi_{\mathbf k_1}\phi_{\mathbf k_2}\phi_{\mathbf k_3}\phi_{\mathbf k_4} \rangle'_{\rm flat{\rm noP}} \times 4 + {\rm permutations}~.
			\end{align} 
		\end{itemize}
	\end{itemize} 
\end{itemize}

\section{Application to Bouncing Cosmology}\label{applicationtobounce} 
In this appendix, we apply the symmetry breaking operator technique to matter bounce cosmology with a general single field content~\cite{Li:2016xjb}. The idea is that the universe underwent a matter contraction phase before the Big Bang~\cite{Wands:1998yp,Finelli:2001sr} (for a review, see~\cite{Cai:2009zp,Brandenberger:2012zb}). We still consider the action of the form $\displaystyle \int(R+2P(X,\phi))$ as in the inflationary scenario. But for the matter bounce, the scale factor is given by
\begin{align}
a(\tau) =  \bigg(\frac{\tau-\tilde \tau_B}{\tau_B - \tilde \tau_B}\bigg)^2 ~.
\end{align}
where $\tau_B$ corresponds to the conformal time at the beginning of the bounce phase and $\tilde \tau_B$ corresponds to the time of the bouncing singularity (if there is no new physics to resolve the singularity; otherwise a non-singular bounce is obtained and $\tilde \tau_B$ is indeed close to the time of the bounce).
The Hubble parameter is
\begin{align}
H = \frac{a'}{a^2} = \frac{2(\tau_B-\tilde \tau_B)^2}{(\tau-\tau_B)^3}~.
\end{align}
The mode function and its derivative is
\begin{equation}
\begin{aligned}
u_k(\tau)=-\frac{iA[1+ic_sk(\tau-\tilde{\tau}_B)]}{2\sqrt{\epsilon c_sk^3}(\tau-\tilde{\tau}_B)^3}e^{-ic_sk(\tau-\tilde{\tau}_B)} 
\end{aligned}
\end{equation}
and its derivative with respect to $\tau$ is 
\begin{equation}
\begin{aligned}
u_k'(\tau)=\frac{iA}{2\sqrt{\epsilon c_sk^3}}\left(\frac{3\left[1+ic_sk(\tau-\tilde{\tau}_B)\right]}{(\tau-\tilde{\tau}_B)^4}-\frac{c_s^2k^2}{(\tau-\tilde{\tau}_B)^2}\right)e^{-ic_sk(\tau-\tilde{\tau}_B)} ~,
\end{aligned}
\end{equation}
where $A=(\tau_B-\tilde{\tau}_B)^2$ and $M_{\rm{Pl}}$ is set to be $1$. Thus the dimensionless power spectrum is 
\begin{equation}
\begin{aligned} 
\Delta_\zeta^2=\frac{1}{12\pi^2c_s(\tau_B-\tilde{\tau}_B)^2}~.
\end{aligned}
\end{equation}

To evaluate the three-point correlation function, the third order action is~\cite{Li:2016xjb}
\begin{equation}\label{action3}
\begin{aligned}
S^{(3)}=\int d\tau\,d^3x\bigg\{&-\frac{a}{H^3}\left[\Sigma\left(1-\frac{1}{c_s^2}\right)+2\lambda\right]\zeta'^3
+a^2\left[\frac{\epsilon}{c_s^4}(\epsilon-3+3c_s^2)-\frac{\epsilon^3}{2}\right]\zeta\zeta'^2
+\frac{a^2\epsilon}{c_s^2}(\epsilon-2s+1-c_s^2)\zeta\partial_i\zeta\partial^i\zeta\\
&-\frac{2a\epsilon}{c_s^2}\zeta'\partial_i\zeta\partial^i\chi
+\frac{a^2\epsilon}{2c_s^2}\frac{d}{d\tau}\left(\frac{\eta}{c_s^2}\right)\zeta^2\zeta'\
+\frac{\epsilon}{2}\zeta(\partial_i\partial_j\chi)(\partial^i\partial^j\chi)
+2f(\zeta)\frac{\delta L}{\delta\zeta}\bigg|_1\bigg\}~,
\end{aligned}
\end{equation}
where  $\chi$ is defined via $\partial^2\chi=a\epsilon\zeta'$, and 
\begin{equation}
\begin{aligned}
f(\zeta)&=\frac{\eta}{4c_s^2}\zeta^2+\frac{1}{c_s^2aH}\zeta\zeta'+\frac{1}{4a^2H^2}\left(-\partial_i\zeta\partial^i\zeta+\partial^{-2}\left[\partial_i\partial_j\left(\partial^i\partial^j\zeta\right)\right]\right)+\frac{1}{2a^2H}\{\partial_i\zeta\partial^i\chi-\partial^{-2}[\partial_i\partial_j(\partial^i\zeta\partial^j\chi)]\}~,\\
\frac{\delta L}{\delta\zeta}\bigg|_1&=a\left(\frac{d}{dt}\partial^2\chi+H\partial^2\chi-\epsilon\partial^2\zeta\right)~.
\end{aligned}
\end{equation}
$\partial^2$ and $\partial^{-2}$ are the Laplacian and the inverse of Laplacian, respectively. If we only consider the case that $c_s$ is nearly constant, that is, $s\approx0$, we have $\lambda/\Sigma\approx$ constant, and the first term can be rewritten as 
\begin{equation}
\begin{aligned}
-\frac{\epsilon}{c_s^2}\left(1-\frac{1}{c_s^2}+\frac{2\lambda}{\Sigma}\right)\frac{a}{H}\zeta'^3~.
\end{aligned}
\end{equation}

The $c_s=1$ case is considered in~\cite{Cai:2009fn}, whereas the general case $c_s\neq 1$ is considered in~\cite{Li:2016xjb}. The last term in this action is removed by performing the field redefinition
\begin{align}
\zeta\rightarrow\tilde\zeta+f(\tilde\zeta)~,
\end{align}
where $\tilde\zeta$ denotes the field after redefinition.

\subsection{Rule to relate external mode function}
For matter bounce, we have the following relation between the mode functions in cosmology and in Minkowski spacetime
\begin{align}
u_k(\tau_B)_{\rm MB} = u_k(c_s(\tau_B-\tilde{\tau}_B))_{\rm flat}\left(-\frac{i\left[1+ic_sk(\tau_B-\tilde{\tau}_B)\right]}{\sqrt{2 \epsilon c_s} k(\tau_B-\tilde{\tau}_B)}\right)~.
\end{align}
For three-point function, we have an extra factor of 
\begin{align}
\frac{i}{2\sqrt{2\epsilon^3c_s^3}k_1k_2k_3(\tau_B-\tilde{\tau}_B)^3}\prod_{i=1}^3\left[1+ic_sk_i(\tau_B-\tilde{\tau}_B)\right]~.
\end{align}

\subsection{Rule to relate Hamiltonian}
\begin{itemize}
	\item The Minkowski Hamiltonian \\
	\begin{itemize}
		\item 	$H_{\phi^3}$, if all fields in it contract with the field on the left hand side, it corresponds to the following integral,
		\begin{align}
		H_{\phi^3} \rightarrow \int_{-\infty}^{c_s(\tau_B-\tilde{\tau}_B)} \frac{\lambda_3}{2 \sqrt{2} \sqrt{k_1k_2k_3} } e^{i k_{123} \tau} d\tau
		\end{align}
	\end{itemize} 
	\item Matter bounce contraction phase Hamiltonian\\
	\begin{itemize}
		\item $H_{\zeta'^3}$, if all fields in it contract with the field on the left hand side, it corresponds to the following integral,
		\begin{align}
		H_{\zeta'^3}\rightarrow-\int_{-\infty}^{\tau_B-\tilde{\tau}_B}d\tau\;\left(1-\frac{1}{c_s^2}+\frac{2\lambda}{\Sigma}\right)\frac{i(\tau_B-\tilde{\tau}_B)^2}{16\sqrt{\epsilon c_s^7k_1^3k_2^3k_3^3}}\tau^5\prod_{i=1}^3\left[\frac{3\left(1-ic_sk_i\tau\right)}{\tau^4}-\frac{c_s^2k_i^2}{\tau^2}\right]e^{ic_sk_{123}\tau} 
		\end{align}
		Comparing with the integral expression of $H_{\phi^3}$, we have
		\begin{align}
		H_{\zeta'^3
		} \rightarrow \left(1-\frac{1}{c_s^2}+\frac{2\lambda}{\Sigma}\right)\frac{i(\tau_B-\tilde{\tau}_B)^2}{4\sqrt{2\epsilon c_s^7}k_1k_2k_3\lambda_3}O_{\zeta'^3}H_{\phi^3}~,
		\end{align}
		where $O_{\zeta'^3}$ is defined as
		\begin{equation}
		\begin{aligned}
		O_{\zeta'^3}(\mathbf{k}_1,\mathbf{k}_2,\mathbf{k}_3)\equiv(ic_s)^7\bigg[&27I^7+27k_{123}I^6+\frac{9}{2}\bigg(2\sum_{i=1}^3k_i^2+3\sum_{i\neq j}k_ik_j\bigg)I^5+9\bigg(\sum_{i\neq j}k_i^2k_j+3k_1k_2k_3\bigg)I^4\\
		&+\frac{3}{2}\bigg(\sum_{i\neq j}k_1^2k_j^2+3\sum_{i\neq j\neq k}k_i^2k_jk_k\bigg)I^3+\frac{3}{2}\bigg(\sum_{i\neq j\neq k}k_i^2k_j^2k_k\bigg)I^2+k_1^2k_2^2k_3^2I\bigg]~.
		\end{aligned}
		\end{equation}
		and $I^n$ means integrating $n$ times with respect to $k_{123}$. Thus the corresponding part of the three-point function is
		\begin{equation}
		\begin{aligned}
		\langle \zeta_{\mathbf k_1}\zeta_{\mathbf k_2}\zeta_{\mathbf k_3} \rangle_{\rm MB} &= \frac{i\prod_{i=1}^3\left[1+ic_sk_i(\tau_B-\tilde{\tau}_B)\right]}{2\sqrt{2\epsilon^3c_s^3}k_1k_2k_3(\tau_B-\tilde{\tau}_B)^3}\left(1-\frac{1}{c_s^2}+\frac{2\lambda}{\Sigma}\right)\frac{i(\tau_B-\tilde{\tau}_B)^2}{4\sqrt{2\epsilon c_s^7}k_1k_2k_3\lambda_3}\\
		&\quad\times O_{\zeta'^3}\langle \phi_{\mathbf k_1}\phi_{\mathbf k_2}\phi_{\mathbf k_3}  \rangle_{\rm flat}\\
		&+\text{5 permutations}
		\end{aligned}
		\end{equation}
		\item $H_{\zeta \zeta'^2}$, if all fields in it contract with the field on the left hand side, it corresponds to the following integral,
		\begin{equation}
		\begin{aligned}
		H_{\zeta\zeta'^2}(k_1,k_2,k_3)\rightarrow&\int_{-\infty}^{\tau_B-\tilde{\tau}_B}d\tau\;\left[\frac{\epsilon}{c_s^4}(\epsilon-3+3c_s^2)-\frac{\epsilon^3}{2}\right]\frac{i(\tau_B-\tilde{\tau}_B)^2}{8\sqrt{\epsilon c_sk_1k_2k_3}^3}\\
		&\times\tau^4\left(\frac{1-ic_sk_1\tau}{\tau^3}\right)\prod_{i=2}^3\left[\frac{3\left(1-ic_sk_i\tau\right)}{\tau^4}-\frac{c_s^2k_i^2}{\tau^2}\right]e^{ic_sk_{123}\tau}
		\end{aligned}
		\end{equation}
		Here $k_1$ corresponds to the momentum of $\zeta$, $k_2$ and $k_3$ corresponds to the momentum of the second or third $\zeta'$, respectively. Comparing with the integral expression of $H_{\phi^3}$, we have
		\begin{equation}
		\begin{aligned}
		H_{\zeta\zeta'^2}(k_1,k_2,k_3)\rightarrow&\left[\frac{\epsilon}{c_s^4}(\epsilon-3+3c_s^2)-\frac{\epsilon^3}{2}\right]\frac{i(\tau_B-\tilde{\tau}_B)^2}{2\sqrt{2\epsilon^3c_s^3}k_1k_2k_3\lambda_3}O_{\zeta\zeta'^2}H_{I3}
		\end{aligned}
		\end{equation}
		where $O_{\zeta\zeta'^2}$ is defined as
		\begin{equation}
		\begin{aligned}
		O_{\zeta\zeta'^2}(\mathbf{k}_1,\mathbf{k}_2,\mathbf{k}_3)&\equiv9I^7+9k_{123}I^6+3(3k_1k_2+3k_2k_3+3k_3k_1+k_2^2+k_3^2)I^5\\
		&\quad\;+3\left(k_1k_2^2+k_1k_3^2+k_2k_3^2+k_2^2k_3+3k_1k_2k_3\right)I^4+\left(3 k_1 k_2^2 k_3 + 3 k_1 k_2 k_3^2 + k_2^2 k_3^2\right)I^3\\
		&\quad\;+\left(k_1 k_2^2 k_3^2\right)I^2~.
		\end{aligned}
		\end{equation}
		Thus the corresponding part of the three-point function is
		\begin{equation}
		\begin{aligned}
		&\langle \zeta_{\mathbf k_1}\zeta_{\mathbf k_2}\zeta_{\mathbf k_3} \rangle_{\rm MB} \\
		=&\frac{i\prod_{i=1}^3\left[1+ic_sk_i(\tau_B-\tilde{\tau}_B)\right]}{2\sqrt{2\epsilon^3c_s^3}k_1k_2k_3(\tau_B-\tilde{\tau}_B)^3}\left[\frac{\epsilon}{c_s^4}(\epsilon-3+3c_s^2)-\frac{\epsilon^3}{2}\right]\frac{i(\tau_B-\tilde{\tau}_B)^2}{2\sqrt{2\epsilon^3c_s^3}k_1k_2k_3\lambda_3}\\
		&\times O_{\zeta\zeta'^2}\langle \phi_{\mathbf k_1}\phi_{\mathbf k_2}\phi_{\mathbf k_3} \rangle_{\rm flat}\\
		&+\text{5 permutations}
		\end{aligned}
		\end{equation}
		
		\item $H_{\zeta(\partial\zeta)^2}$, if all fields in it contract with the field on the left hand side, it corresponds to the following integral,
		\begin{equation}
		\begin{aligned}
		H_{\zeta(\partial\zeta)^2}(k_1,k_2,k_3)\rightarrow&\int_{-\infty}^{\tau_B-\tilde{\tau}_B}d\tau\;\frac{i(\epsilon+1-c_s^2)(\tau_B-\tilde{\tau}_B)^2}{8\sqrt{\epsilon c_s^7k_1^3k_2^3k_3^3}}\mathbf{k_2}\cdot\mathbf{k_3}\tau^4\prod_{i=1}^3\left[\frac{3\left(1-ic_sk_i\tau\right)}{\tau^4}-\frac{c_s^2k_i^2}{\tau^2}\right]e^{ic_sk_{123}\tau}
		\end{aligned}
		\end{equation}
		Here $k_1$ corresponds to the momentum of the first $\zeta$, $k_2$ and $k_3$ corresponds to the momentum of the second or third $\zeta$, respectively. Comparing with the integral expression of $H_{\phi^3}$, we have
		\begin{equation}
		\begin{aligned}
		H_{\zeta(\partial\zeta)^2}(k_1,k_2,k_3)\rightarrow&\frac{i(\epsilon+1-c_s^2)(\tau_B-\tilde{\tau}_B)^2}{2\sqrt{2\epsilon c_s^7}k_1k_2k_3\lambda_3}O_{\zeta(\partial\zeta)^2}H_{\phi^3}
		\end{aligned}
		\end{equation}
		where ${O}_{\zeta(\partial\zeta)^2}$ is defined as
		\begin{equation}
		\begin{aligned}
		{O}_{\zeta(\partial\zeta)^2}(\mathbf{k}_1,\mathbf{k}_2,\mathbf{k}_3)&\equiv(ic_s)^5\bigg(I^5+k_{123}I^4+\frac{1}{2}\sum_{i\neq j}k_ik_jI^3+k_1k_2k_3I^2\bigg)~.
		\end{aligned}
		\end{equation}
		
		Thus the corresponding part of the three-point function is
		\begin{equation}
		\begin{aligned}
		\langle \zeta_{\mathbf k_1}\zeta_{\mathbf k_2}\zeta_{\mathbf k_3} \rangle_{\rm MB} =&\frac{i\prod_{i=1}^3\left[1+ic_sk_i(\tau_B-\tilde{\tau}_B)\right]}{2\sqrt{2\epsilon^3c_s^3}k_1k_2k_3(\tau_B-\tilde{\tau}_B)^3}\frac{i(\epsilon+1-c_s^2)(\tau_B-\tilde{\tau}_B)^2}{2\sqrt{2\epsilon c_s^7}k_1k_2k_3\lambda_3}\\
		&\times O_{\zeta(\partial\zeta)^2}\langle \phi_{\mathbf k_1}\phi_{\mathbf k_2}\phi_{\mathbf k_3} \rangle_{\rm flat}\\
		&+\text{5 permutations}
		\end{aligned}
		\end{equation}
		
		\item $H_{\zeta' \partial\zeta \partial\chi}$, if all fields in it contract with the field on the left hand side, it corresponds to the following integral,
		\begin{equation}
		\begin{aligned}
		H_{\zeta' \partial\zeta \partial\chi}(k_1,k_2,k_3)\rightarrow&-\int_{-\infty}^{\tau_B-\tilde{\tau}_B}d\tau\;\frac{i\sqrt{\epsilon}(\tau_B-\tilde{\tau}_B)^2}{4\sqrt{c_s^7k_1^3k_2^3k_3^3}}\frac{\mathbf{k_2}\cdot\mathbf{k_3}}{k_3^2}\tau^4\left(\frac{1-ic_sk_2\tau}{\tau^3}\right)\prod_{i=1,3}\left[\frac{3\left(1-ic_sk_i\tau\right)}{\tau^4}-\frac{c_s^2k_i^2}{\tau^2}\right]e^{ic_sk_{123}\tau}
		\end{aligned}
		\end{equation}
		Here $k_1$ and $k_2$ corresponds to the momentum of the first or second $\zeta$, $k_3$ corresponds to the momentum of $\chi$, respectively. Comparing with the integral expression of $H_{\phi^3}$, we have
		\begin{equation}
		\begin{aligned}
		H_{\zeta' \partial\zeta \partial\chi}(k_1,k_2,k_3)\rightarrow&-\frac{i\sqrt{\epsilon}(\tau_B-\tilde{\tau}_B)^2}{\sqrt{2c_s^7}k_1k_2k_3\lambda_3}O_{\zeta' \partial\zeta \partial\chi}H_{\phi^3}
		\end{aligned}
		\end{equation}
		where $O_{\zeta' \partial\zeta \partial\chi}$ is defined as
		\begin{equation}
		\begin{aligned}
		O_{\zeta'\partial\zeta\partial\chi}(\mathbf{k}_1,\mathbf{k}_2,\mathbf{k}_3)&\equiv\frac{\mathbf{k}_2\cdot\mathbf{k}_3}{k_3^2}{O}_{\zeta\zeta'^2}(\mathbf{k}_1,\mathbf{k}_2,\mathbf{k}_3)~.
		\end{aligned}
		\end{equation}
		Thus the corresponding part of the three-point function is
		\begin{equation}
		\begin{aligned}
		\langle \zeta_{\mathbf k_1}\zeta_{\mathbf k_2}\zeta_{\mathbf k_3} \rangle_{\rm MB} =&-\frac{i\prod_{i=1}^3\left[1+ic_sk_i(\tau_B-\tilde{\tau}_B)\right]}{2\sqrt{2\epsilon^3c_s^3}k_1k_2k_3(\tau_B-\tilde{\tau}_B)^3}\frac{i\sqrt{\epsilon}(\tau_B-\tilde{\tau}_B)^2}{\sqrt{2c_s^7}k_1k_2k_3\lambda_3}\\
		&\times O_{\zeta' \partial\zeta \partial\chi}\langle \phi_{\mathbf k_1}\phi_{\mathbf k_2}\phi_{\mathbf k_3}  \rangle_{\rm flat}\\
		&+\text{5 permutations}
		\end{aligned}
		\end{equation}
		
		\item $H_{\zeta(\partial_i\partial_j\chi)^2}$, if all fields in it contract with the field on the left hand side, it corresponds to the following integral,
		\begin{equation}
		\begin{aligned}
		H_{\zeta(\partial_i\partial_j\chi)^2}(k_1,k_2,k_3)\rightarrow&\int_{-\infty}^{\tau_B-\tilde{\tau}_B}d\tau\;\frac{i\sqrt{\epsilon^3}(\tau_B-\tilde{\tau}_B)^2}{8\sqrt{c_s^3k_1^3k_2^3k_3^3}}\left(\frac{\mathbf{k_2}\cdot\mathbf{k_3}}{k_2k_3}\right)^2\tau^4\left(\frac{1-ic_sk_1\tau}{\tau^3}\right)\\
		&\quad\;\prod_{i=2}^3\left[\frac{3\left(1-ic_sk_i\tau\right)}{\tau^4}-\frac{c_s^2k_i^2}{\tau^2}\right]e^{ic_sk_{123}\tau}
		\end{aligned}
		\end{equation}
		Here $k_1$ corresponds to the momentum of $\zeta$, $k_2$ and $k_3$ corresponds to the momentum of the second or third $\chi$, respectively. Comparing with the integral expression of $H_{\phi^3}$, we have
		\begin{equation}
		\begin{aligned}
		H_{\zeta(\partial_i\partial_j\chi)^2}(k_1,k_2,k_3)\rightarrow&\frac{i\sqrt{\epsilon^3}(\tau_B-\tilde{\tau}_B)^2}{2\sqrt{2c_s^3}k_1k_2k_3\lambda_3}O_{\zeta(\partial_i\partial_j\chi)^2}H_{\phi^3}
		\end{aligned}
		\end{equation}
		where $O_{\zeta(\partial_i\partial_j\chi)^2}$ is defined as
		\begin{equation}
		\begin{aligned}
		O_{\zeta(\partial_i\partial_j\chi)^2}(\mathbf{k}_1,\mathbf{k}_2,\mathbf{k}_3)&\equiv\left(\frac{\mathbf{k}_2\cdot\mathbf{k}_3}{k_2k_3}\right)^2\mathcal{O}_{\zeta\zeta'^2}(\mathbf{k}_1,\mathbf{k}_2,\mathbf{k}_3)~.
		\end{aligned}
		\end{equation}
		Thus the corresponding part of the three-point function is
		\begin{equation}
		\begin{aligned}
		\langle \zeta_{\mathbf k_1}\zeta_{\mathbf k_2}\zeta_{\mathbf k_3} \rangle_{\rm MB} =&-\frac{i\prod_{i=1}^3\left[1+ic_sk_i(\tau_B-\tilde{\tau}_B)\right]}{2\sqrt{2\epsilon^3c_s^3}k_1k_2k_3(\tau_B-\tilde{\tau}_B)^3}\frac{i\sqrt{\epsilon^3}(\tau_B-\tilde{\tau}_B)^2}{2\sqrt{2c_s^3}k_1k_2k_3\lambda_3}\\
		&\times O_{\zeta(\partial_i\partial_j\chi)^2}\langle \phi_{\mathbf k_1}\phi_{\mathbf k_2}\phi_{\mathbf k_3} \rangle_{\rm flat}\\
		&+\text{5 permutations}~.
		\end{aligned}
		\end{equation}
		
		\item Secondary contributions: The contribution from the term
		\begin{equation}
		\frac{a^2\epsilon}{2c_s^2}\frac{d}{d\tau}\left(\frac{\eta}{c_s^2}\right)\zeta^2\zeta'\nonumber
		\end{equation}
		in equation\ \eqref{action3} is exactly zero since $\eta=0$ during the matter contraction.
		We can also neglect the contribution from the term
		\begin{equation}
		\frac{a\epsilon}{c_\mathrm{s}^2}(\epsilon-2s+1-c_\mathrm{s}^2)\zeta(\partial\zeta)^2 \nonumber
		\end{equation}
		since the leading order term of the resulting three-point function is proportional to $c_\mathrm{s}^2k_i^2(\tau_B-\tilde{\tau}_B)^2$,
		which means that this term is suppressed outside the sound horizon.
		
	\end{itemize}
\end{itemize}


\begin{thebibliography}{999} 

\bibitem{Arkani-Hamed:2015bza} 
N.~Arkani-Hamed and J.~Maldacena,
``Cosmological Collider Physics,''
arXiv:1503.08043 [hep-th].

\bibitem{Chen:2009we} 
X.~Chen and Y.~Wang,
``Large non-Gaussianities with Intermediate Shapes from Quasi-Single Field Inflation,''
Phys.\ Rev.\ D {\bf 81}, 063511 (2010)
[arXiv:0909.0496 [astro-ph.CO]].

\bibitem{Chen:2009zp} 
X.~Chen and Y.~Wang,
``Quasi-Single Field Inflation and Non-Gaussianities,''
JCAP {\bf 1004}, 027 (2010)
[arXiv:0911.3380 [hep-th]].

\bibitem{Baumann:2011nk} 
  D.~Baumann and D.~Green,
  ``Signatures of Supersymmetry from the Early Universe,''
  Phys.\ Rev.\ D {\bf 85}, 103520 (2012)
  [arXiv:1109.0292 [hep-th]].
  
\bibitem{Assassi:2012zq} 
V.~Assassi, D.~Baumann and D.~Green,
``On Soft Limits of Inflationary Correlation Functions,''
JCAP {\bf 1211}, 047 (2012)
[arXiv:1204.4207 [hep-th]].

\bibitem{Noumi:2012vr} 
T.~Noumi, M.~Yamaguchi and D.~Yokoyama,
``Effective field theory approach to quasi-single field inflation and effects of heavy fields,''
JHEP {\bf 1306}, 051 (2013)
[arXiv:1211.1624 [hep-th]].

\bibitem{Arkani-Hamed:2017fdk} 
N.~Arkani-Hamed, P.~Benincasa and A.~Postnikov,
``Cosmological Polytopes and the Wavefunction of the Universe,''
arXiv:1709.02813 [hep-th].

\bibitem{Britto:2005fq} 
R.~Britto, F.~Cachazo, B.~Feng and E.~Witten,
``Direct proof of tree-level recursion relation in Yang-Mills theory,''
Phys.\ Rev.\ Lett.\  {\bf 94}, 181602 (2005)
[hep-th/0501052].

\bibitem{ArkaniHamed:2010kv} 
N.~Arkani-Hamed, J.~L.~Bourjaily, F.~Cachazo, S.~Caron-Huot and J.~Trnka,
``The All-Loop Integrand For Scattering Amplitudes in Planar N=4 SYM,''
JHEP {\bf 1101}, 041 (2011)
[arXiv:1008.2958 [hep-th]].

\bibitem{ArkaniHamed:2012nw}
N.~Arkani-Hamed, J.~L.~Bourjaily, F.~Cachazo, A.~B.~Goncharov, A.~Postnikov and J.~Trnka,
``Grassmannian Geometry of Scattering Amplitudes,''
arXiv:1212.5605 [hep-th].

\bibitem{Benincasa:2015zna} 
P.~Benincasa,
``On-shell diagrammatics and the perturbative structure of planar gauge theories,''
arXiv:1510.03642 [hep-th].

\bibitem{Weinberg:2005vy} 
  S.~Weinberg,
  ``Quantum contributions to cosmological correlations,''
  Phys.\ Rev.\ D {\bf 72}, 043514 (2005)
  [hep-th/0506236].

\bibitem{Chen:2010xka} 
  X.~Chen,
  ``Primordial Non-Gaussianities from Inflation Models,''
  Adv.\ Astron.\  {\bf 2010}, 638979 (2010)
  [arXiv:1002.1416 [astro-ph.CO]].

\bibitem{Wang:2013eqj} 
  Y.~Wang,
  ``Inflation, Cosmic Perturbations and Non-Gaussianities,''
  Commun.\ Theor.\ Phys.\  {\bf 62}, 109 (2014)
  [arXiv:1303.1523 [hep-th]].

\bibitem{Chen:2006nt} 
  X.~Chen, M.~x.~Huang, S.~Kachru and G.~Shiu,
  ``Observational signatures and non-Gaussianities of general single field inflation,''
  JCAP {\bf 0701}, 002 (2007)
  [hep-th/0605045].

\bibitem{Strominger:2001pn} 
A.~Strominger,
``The dS / CFT correspondence,''
JHEP {\bf 0110}, 034 (2001)
[hep-th/0106113].

\bibitem{Strominger:2001gp} 
A.~Strominger,
``Inflation and the dS / CFT correspondence,''
JHEP {\bf 0111}, 049 (2001)
[hep-th/0110087].

\bibitem{McFadden:2010na} 
P.~McFadden and K.~Skenderis,
``The Holographic Universe,''
J.\ Phys.\ Conf.\ Ser.\  {\bf 222}, 012007 (2010)
[arXiv:1001.2007 [hep-th]].

\bibitem{McFadden:2010vh} 
P.~McFadden and K.~Skenderis,
``Holographic Non-Gaussianity,''
JCAP {\bf 1105}, 013 (2011)
[arXiv:1011.0452 [hep-th]].

\bibitem{Isono:2016yyj}
H.~Isono, T.~Noumi, G.~Shiu, S.~S.~C.~Wong and S.~Zhou,
``Holographic non-Gaussianities in general single-field inflation,''
JHEP {\bf 1612} (2016) 028
[arXiv:1610.01258 [hep-th]].

\bibitem{Piao:2011bz} 
Y.~S.~Piao,
arXiv:1112.3737 [hep-th].

\bibitem{Maldacena:2002vr} 
J.~M.~Maldacena,
``Non-Gaussian features of primordial fluctuations in single field inflationary models,''
JHEP {\bf 0305}, 013 (2003)
[astro-ph/0210603].

\bibitem{Acquaviva:2002ud} 
V.~Acquaviva, N.~Bartolo, S.~Matarrese and A.~Riotto,
``Second order cosmological perturbations from inflation,''
Nucl.\ Phys.\ B {\bf 667}, 119 (2003)
[astro-ph/0209156].

\bibitem{ArmendarizPicon:1999rj} 
C.~Armendariz-Picon, T.~Damour and V.~F.~Mukhanov,
``k - inflation,''
Phys.\ Lett.\ B {\bf 458}, 209 (1999)
[hep-th/9904075].

\bibitem{Alishahiha:2004eh} 
M.~Alishahiha, E.~Silverstein and D.~Tong,
``DBI in the sky,''
Phys.\ Rev.\ D {\bf 70}, 123505 (2004)
[hep-th/0404084].

\bibitem{Seery:2005wm} 
D.~Seery and J.~E.~Lidsey,
``Primordial non-Gaussianities in single field inflation,''
JCAP {\bf 0506}, 003 (2005)
[astro-ph/0503692].

\bibitem{Hartle:1983ai} 
J.~B.~Hartle and S.~W.~Hawking,
``Wave Function of the Universe,''
Phys.\ Rev.\ D {\bf 28}, 2960 (1983).

\bibitem{Hertog:2011ky} 
T.~Hertog and J.~Hartle,
``Holographic No-Boundary Measure,''
JHEP {\bf 1205}, 095 (2012)
[arXiv:1111.6090 [hep-th]].

\bibitem{Bunch:1978yq} 
T.~S.~Bunch and P.~C.~W.~Davies,
``Quantum Field Theory in de Sitter Space: Renormalization by Point Splitting,''
Proc.\ Roy.\ Soc.\ Lond.\ A {\bf 360}, 117 (1978).

\bibitem{Bzowski:2013sza} 
A.~Bzowski, P.~McFadden and K.~Skenderis,
``Implications of conformal invariance in momentum space,''
JHEP {\bf 1403}, 111 (2014)
[arXiv:1304.7760 [hep-th]].

\bibitem{Bzowski:2015pba} 
A.~Bzowski, P.~McFadden and K.~Skenderis,
``Scalar 3-point functions in CFT: renormalisation, beta functions and anomalies,''
JHEP {\bf 1603}, 066 (2016)
[arXiv:1510.08442 [hep-th]].

\bibitem{Bzowski:2015yxv} 
A.~Bzowski, P.~McFadden and K.~Skenderis,
``Evaluation of conformal integrals,''
JHEP {\bf 1602}, 068 (2016)
[arXiv:1511.02357 [hep-th]].

\bibitem{Chernikov:1968zm} 
N.~A.~Chernikov and E.~A.~Tagirov,
``Quantum theory of scalar fields in de Sitter space-time,''
Ann.\ Inst.\ H.\ Poincare Phys.\ Theor.\ A {\bf 9}, 109 (1968).

\bibitem{Seery:2008ax} 
D.~Seery, M.~S.~Sloth and F.~Vernizzi,
``Inflationary trispectrum from graviton exchange,''
JCAP {\bf 0903}, 018 (2009)
[arXiv:0811.3934 [astro-ph]].


\bibitem{Hiroshi}
H.~Isono, T.~Noumi and G. Shiu, to appear


\bibitem{Adams:2006sv} 
A.~Adams, N.~Arkani-Hamed, S.~Dubovsky, A.~Nicolis and R.~Rattazzi,
``Causality, analyticity and an IR obstruction to UV completion,''
JHEP {\bf 0610}, 014 (2006)
[hep-th/0602178].

\bibitem{Baumann:2015nta} 
D.~Baumann, D.~Green, H.~Lee and R.~A.~Porto,
``Signs of Analyticity in Single-Field Inflation,''
Phys.\ Rev.\ D {\bf 93}, no. 2, 023523 (2016)
[arXiv:1502.07304 [hep-th]].

\bibitem{Raju:2012zr} 
S.~Raju,
``New Recursion Relations and a Minkowski Limit for AdS/CFT Correlators,''
Phys.\ Rev.\ D {\bf 85}, 126009 (2012)
[arXiv:1201.6449 [hep-th]].

\bibitem{Arkani-Hamed:2013jha} 
N.~Arkani-Hamed and J.~Trnka,
``The Amplituhedron,''
JHEP {\bf 1410}, 030 (2014)
[arXiv:1312.2007 [hep-th]].

\bibitem{Arkani-Hamed:2013kca} 
N.~Arkani-Hamed and J.~Trnka,
``Into the Amplituhedron,''
JHEP {\bf 1412}, 182 (2014)
[arXiv:1312.7878 [hep-th]].

\bibitem{Arkani-Hamed:2017vfh} 
N.~Arkani-Hamed, H.~Thomas and J.~Trnka,
``Unwinding the Amplituhedron in Binary,''
JHEP {\bf 1801}, 016 (2018)
[arXiv:1704.05069 [hep-th]].

\bibitem{Chen:2009bc}
X.~Chen, B.~Hu, M.~x.~Huang, G.~Shiu and Y.~Wang,
``Large Primordial Trispectra in General Single Field Inflation,''
JCAP {\bf 0908} (2009) 008
[arXiv:0905.3494 [astro-ph.CO]].

\bibitem{Arroja:2009pd} 
F.~Arroja, S.~Mizuno, K.~Koyama and T.~Tanaka,
``On the full trispectrum in single field DBI-inflation,''
Phys.\ Rev.\ D {\bf 80}, 043527 (2009)
[arXiv:0905.3641 [hep-th]].

\bibitem{Huang:2006eha} 
X.~Chen, M.~x.~Huang and G.~Shiu,
``The Inflationary Trispectrum for Models with Large Non-Gaussianities,''
Phys.\ Rev.\ D {\bf 74}, 121301 (2006)
[hep-th/0610235].

\bibitem{Arroja:2008ga} 
F.~Arroja and K.~Koyama,
``Non-gaussianity from the trispectrum in general single field inflation,''
Phys.\ Rev.\ D {\bf 77}, 083517 (2008)
[arXiv:0802.1167 [hep-th]].

\bibitem{Nandan:2016ohb} 
  D.~Nandan, J.~Plefka and W.~Wormsbecher,
  ``Collinear limits beyond the leading order from the scattering equations,''
  JHEP {\bf 1702}, 038 (2017)
  [arXiv:1608.04730 [hep-th]].

\bibitem{Jiang:2015hfa} 
  H.~Jiang and Y.~Wang,
  ``Towards the physical vacuum of cosmic inflation,''
  Phys.\ Lett.\ B {\bf 760}, 202 (2016)
  [arXiv:1507.05193 [hep-th]].
  
\bibitem{Jiang:2016nok} 
  H.~Jiang, Y.~Wang and S.~Zhou,
  ``On the initial condition of inflationary fluctuations,''
  JCAP {\bf 1604}, no. 04, 041 (2016)
  [arXiv:1601.01179 [hep-th]].

\bibitem{Cai:2009fn} 
Y.~F.~Cai, W.~Xue, R.~Brandenberger and X.~Zhang,
``Non-Gaussianity in a Matter Bounce,''
JCAP {\bf 0905}, 011 (2009)
[arXiv:0903.0631 [astro-ph.CO]].

\bibitem{Li:2016xjb} 
Y.~B.~Li, J.~Quintin, D.~G.~Wang and Y.~F.~Cai,
``Matter bounce cosmology with a generalized single field: non-Gaussianity and an extended no-go theorem,''
JCAP {\bf 1703}, no. 03, 031 (2017)
[arXiv:1612.02036 [hep-th]].

\bibitem{Wands:1998yp} 
D.~Wands,
``Duality invariance of cosmological perturbation spectra,''
Phys.\ Rev.\ D {\bf 60}, 023507 (1999)
[gr-qc/9809062].

\bibitem{Finelli:2001sr} 
F.~Finelli and R.~Brandenberger,
``On the generation of a scale invariant spectrum of adiabatic fluctuations in cosmological models with a contracting phase,''
Phys.\ Rev.\ D {\bf 65}, 103522 (2002)
[hep-th/0112249].

\bibitem{Cai:2009zp} 
  Y.~F.~Cai, E.~N.~Saridakis, M.~R.~Setare and J.~Q.~Xia,
  ``Quintom Cosmology: Theoretical implications and observations,''
  Phys.\ Rept.\  {\bf 493}, 1 (2010)
  [arXiv:0909.2776 [hep-th]].

\bibitem{Brandenberger:2012zb} 
R.~H.~Brandenberger,
``The Matter Bounce Alternative to Inflationary Cosmology,''
arXiv:1206.4196 [astro-ph.CO].

\end{thebibliography}
\end{document}